\documentclass[useAMS,usenatbib,usegraphicx]{mn2e}

\def\apj{ApJ}
\def\apjl{ApJ}
\def\colwidth{84mm}

\title[Repeating microlensing events]
{Repeating microlensing events in the OGLE data}
\author[J. Skowron, et al.]{
J. Skowron$^{1}$\thanks{
E-mail: (JS) jskowron@astrouw.edu.pl; 
({\L}W) wyrzykow@ast.cam.ac.uk; 
(SM) shude.mao@manchester.ac.uk;
(MJ) mj@astrouw.edu.pl},
{\L}. Wyrzykowski$^{1,2}$\footnotemark[1],
S. Mao$^{3}$\footnotemark[1],
M. Jaroszy{\'n}ski$^{1}$\footnotemark[1]\\
$^{1}$Warsaw University Astronomical Observatory, 
Al. Ujazdowskie 4, 00-478 Warszawa, Poland\\
$^{2}$Institute of Astronomy, University of Cambridge, 
Madingley Road, CB3 0HA Cambridge, UK\\
$^{3}$Alan Turing Building, Jodrell Bank Centre for Astrophysics, 
The University of Manchester, M13 9PL Manchester, UK}
\begin{document}

\date{Accepted 2008 November 13. Received 2008 November 13; in original form 2008 September 10}

\pagerange{\pageref{firstpage}--\pageref{lastpage}} \pubyear{2008}

\maketitle

\label{firstpage}

\begin{abstract}
Microlensing events are usually selected among single-peaked 
non-repeating light curves in order to avoid
confusion with variable stars. However, a microlensing event may exhibit 
a second microlensing brightening episode when the source or/and 
the lens is a binary system. 
A careful analysis of these repeating events provides an independent way to
study the statistics of wide binary stars and to detect extrasolar
planets. Previous theoretical studies predicted that 
{0.5 - 2 \%} of events should repeat due to wide binary lenses.  
We present a systematic search for such events in about 4000 light curves of microlensing candidates detected
by the Optical Gravitational Lensing Experiment (OGLE) towards the Galactic
Bulge from 1992 to 2007. The search reveals a total of 19 repeating candidates, with 6 clearly due to a wide binary lens. 
As a by-product we find that 64 events  ($\sim 2\%$ of the total
OGLE-III sample) have been miss-classified as microlensing; these miss-classified
events are mostly nova or other types of eruptive stars. The number and
importance of repeating events will increase considerably when the
next-generation wide-field microlensing experiments become fully operational in the future.
\end{abstract}


\section{Introduction}

Microlensing events are usually assumed to be single-peaked and constant in 
the baseline before and after the event. Any additional brightening
episode is usually interpreted as an indication of some type of variability other than microlensing.
It is a conservative assumption adopted at the beginning of microlensing surveys in the early 1990s. Even now it is still imposed
in microlensing searches in very crowded fields heavily contaminated with eruptive variable 
stars, e.g.  towards Magellanic Clouds or M31.  

However, microlensing is now well established and many events are
routinely discovered in real-time.
It is well-known that in principle a small fraction of microlensing events
can repeat.  By repetition we mean a second brightening episode well after
the first one such that the star has (largely) returned to its baseline. A variety of
scenarios can cause repetitions,
{for} example,
when the lensed source is a binary wide enough for the lens to magnify both
{components} separately, one after another.
The binarity of source does not have to be physical, as the lens can
sequentially  magnify two unrelated stars blended in one seeing disk. 
Another mechanism that can produce two independent peaks in the light curve
is 
{a lens consisting of two gravitationally bound components with large
separation} magnifying a single source. 
{The lensing of a physical binary by
  another physical binary can result in diverse light curve
  shapes, which in some cases may mimic repeating events. This 
  rather complicated scenario is not pursued in this analysis of limited
  number of candidate events.}
One can even imagine two independent {masses acting as lenses}.
{This is}, however, expected to occur with a much smaller probability
since the microlensing optical depth for a single event is already low ($\sim
10^{-6}$), and thus we do not study {such a scenario} in this paper.

{\citet{Griest92} investigated the effects of the binarity of
  sources on the properties of light curves produced by a single
  lens. They concluded that,  depending on the adopted population
  of the binary population and the lens mass function, 
  the cases with double
  peaks and asymmetric light curves (easily distinguishable from
  single source events) comprise 1\% to 3\% of all 
  events. 
}

\citet{1996ApJ...457...93D} studied the possibility of repeating
  microlensing due to wide binary lenses and predicted  
that 0.5-2 per cent of observed microlensing curves should exhibit apparent
repetitions. The fraction is lower than that predicted for 
``ordinary'' binary events with smaller separations (\citealt{MP91}). 
To date, hundreds of ordinary binary microlensing events have been discovered
  (\citealt{Alc00, Coll04, Jar04, Jar06, 2007AcA....57..281S}), 
but only one ``repeating'' binary event was found  
{(OGLE-2003-BLG-291) and was} studied in detail by
  \citet{2005AcA....55..159J}.  

Thanks to microlensing surveys such as 
the Optical Gravitational Lensing Experiment (OGLE, \citealt{2003AcA....53..291U} and Microlensing Observations in
Astrophysics (MOA, \citealt{MOA}),
several thousand microlensing events have been detected in real-time\footnote{http://ogle.astrouw.edu.pl/ogle3/ews/ews.html; http://www.phys.canterbury.ac.nz/moa/microlensing\_alerts.html}. Among these, we would expect tens of events caused by wide binaries. 

Detection and studies of these events are interesting for two important reasons. First, these events provide an independent way
of studying the statistical properties of the binary star population, including the distribution of mass ratios and separations.
{Since a large fraction of lens population is expected to be composed
  of low-mass ($\sim 0.3M_\odot$) main sequence} stars, the detection of
the binary companions around these faint stars would be difficult using
spectroscopic methods (e.g. \citealt{Fisher05}), {because of their
  typically long periods, small accelerations and slow radial velocity variations.}
The novel aspect of the microlensing method is that we can infer the
mass ratios in the time domain {simply from the ratio of the timescales in the
two brightening episodes (see \S\ref{sec:results}) rather than using
spectroscopy} \citep{1996ApJ...457...93D}. 
Second, repeating microlensing events can in principle be used to detect planetary companions \citep{stefano99}; this would be an extension of
the methods used currently by the survey and followup teams for successful detections of 
extrasolar planets (e.g. \citealt{Bond04, Uda05, Bea06, Gou06, Gaudi08}).

In this work we present the results of a systematic search for repeating events in the available data on all microlensing 
candidates detected by {the OGLE team. The paper is organised as follows.}
Sections \ref{data} and \ref{search} describe observational data and our search procedure. 
In Section \ref{modelling} we focus on different microlensing models and algorithms to fit the light curves of repeating candidates.
The major results of our search are presented in Section \ref{results}, and
we summarise and further discuss our results in Section \ref{concl}.

\section{Observational data}
\label{data}

In this study we use data acquired by the first, second and third
phases of OGLE. The readers are referred to 
\citet{2003AcA....53..291U} for more technical details about these
phases of the project. We gather all the microlensing events detected by the OGLE Early 
Warning System (EWS), by OGLE-II (1998-2000, \citealt{1994AcA....44..227U}) and by OGLE-III (2001-2007, \citealt{2003AcA....53..291U}). 
We also include events found independently by \citet{2001AcA....51..175W} in the OGLE-II  1997-1999 data 
and by \citet{PhD} and \citet{2006AcA....56..145W} in the OGLE-III 2001-2005 data. 
In total there are 4120 unique microlensing candidate events.
{The systematic checking of a possible later (or an overlooked earlier)
episode of variability of sources belonging to EWS catalogs is not
routinely done, and this task is undertaken here. Notice that
events with closely separated peaks may be removed from the candidate list, and thus we focus on ``repeating''
events where the star has (largely) returned to its baseline between the episodes
since such events will not be rejected by the EWS.}
%

For each event, we check if any additional observational data are
available in the OGLE-I (1992-1995), OGLE-II (1996-2000) and OGLE-III (2001-2007)
databases. The maximum available time span is therefore 15 years.
152 microlensing events were observed continuously during all three phases of the
OGLE  project (15 years). 
About 1200 events were observed continuously for 10 years and about 2300 for 5 years.

\section{Searching for repeating events}
\label{search}

For each of the 4120 light curves, we search for the presence of two
or more magnification episodes in the whole time span by
visual inspection and with a semi-automated algorithm. We discuss these two approaches in turn.

\subsection{Visual inspection}
\label{visual}

The main criterion used in this subjective analysis is the presence of
a significant bump before or after the main brightening episode. 
{We look for events, where the brightenings are separated by a return
  of the light curve to the baseline.}
Our systematic inspection of all the 4120 light curves has revealed 
13 candidates of repeating events.

In addition, a small number of objects have been found
{among the microlensing candidates being other types of brightenings.
They are present only in among events listed by the EWS because of the 
nature of the EWS detection algorithm, which detects significant and 
continuous brightening, but does not perform any microlensing model fitting.}
The main contamination is from nova-like
outbursts and from  other variables, mostly eruptive stars.  The long time
span allows a better classification  of the events, e.g. by identifying
later outbursts or another period in oscillations.  In addition most of
these stars exhibit some asymmetry in their  brightening episodes.

The largest and most uniform sample is from the EWS in the OGLE-III phase
(until  September 2007). 
Out of 3159 microlensing candidates in this sample, 64 ($\sim 2\% $)
turn out to be 
{intrinsically} variable stars; 24 of these show a behaviour  
similar to dwarf novae stars with multiple, short-time outbursts. 
Furthermore, 52 events ($\sim 1.6\% $) have duplicate entries since they
were ``discovered" twice in overlapping adjacent fields.

\subsection{Semi-automated search algorithm}
\label{automatic}

The main goal for developing an automated algorithm is to identify repeating 
events in which the second magnification episode is considerably smaller than 
the first one, as such configuration would be easier to miss visually.
Because of the contamination by variable stars and caustic crossing
events we were only able to construct a semi-automated algorithm, which
still required some human supervision. 

In the first step we detect the main microlensing episode and fit it with the 
\citet{1986ApJ...304....1P} model. {The portion of this
  model light curve which is above $3\sigma$ level from the baseline
  defines the duration of the brightening episode, and the corresponding
  data points are removed from the light curve. The fit of the
  Paczy{\'n}ski model does not have to be perfect, since it only serves to
  mark the brightening episode, and our experience shows that the same
  algorithm can also be successfully applied to binary lensing
  events.}

A constant light curve 
model and the Paczy{\'n}ski model are then fit to the remaining data, 
yielding two $\chi^2$'s: $\chi^2_{\rm 2nd, const}$, and $\chi^2_{\rm 2nd,
  Pac}$. 
If a repeating event is present, $\chi^2_{\rm 2nd, Pac}$  is expected
to be substantially smaller than $\chi^2_{\rm 2nd, const}$.

To ensure the second fit is reliable, we calculate the number of data 
points in the second magnification episode, $n_{\rm 2nd}$, and 
reject all events with $n_{\rm 2nd} < 3$ or $n_{\rm 2nd} > N/2$, where $N$ 
is the total number of data points. We then construct the following statistic:
\begin{equation}
 s = \frac{ | \chi^2_{\rm 2nd, Pac} - \chi^2_{\rm 2nd, const} | }
                              {\chi^2_{\rm 2nd, Pac}},
\end{equation}
and choose events with $s>0.2$. {This ad hoc criterion typically
corresponds to a fit improvement in $\Delta\chi^2$ of several tens, which
is high enough to avoid false identifications (cf. \S 5.1.2). On the other
hand all 13 events found by visual inspection are
recovered.} 

A total of 193 events have passed the described criteria. All of them
have been examined visually 
giving  6 new candidates, and bringing
the total number of candidates for repeating events to 19 (see Table
\ref{table:events}). 
{6 new candidates were overlooked during the visual inspection due to
their small amplitude of the secondary peak.}

3 out of 19 of our candidates were found in previous studies:
the connection between two EWS events OGLE-1999-BUL-42 and OGLE-2003-BLG-220 
was found by \citet{Klimentowski}; OGLE174828.55-221639.9 was found by
\citet{PhD} and OGLE-2003-BLG-291 was described in detail by \citet{2005AcA....55..159J}.

\section{Modeling}
\label{modelling}
\label{sec:model}

All 19 events were fitted with three different models: binary source,
binary lens and approximate wide binary lens models. 

We model a binary lens following \citet{2007AcA....57..281S}. In total
there are seven parameters. The two point lenses are described by the mass
ratio ($q$) and separation ($d$) in units of the 
Einstein radius ($r_E$). The microlensing geometry is described by  
the impact parameter ($b$), the angle between the source trajectory and the
projected binary  
axis ($\beta$ in degrees), Einstein-radius crossing time ($t_E$ in days), 
the time of the closest approach to the center of mass of the binary
($t_0$), and the fraction of light contributed to the total blended flux by
the lensed source, $f$ ($f=1$ indicates no blending). 
The event baseline magnitude ($I_0$) is measured separately.
A point-like source is assumed in the fitting. 
In some cases in order to better fit the observational data we further
introduce the parallax motion of the Earth   
(with a parallax scale $\pi_E$, defined as $1{\rm a.u.}/\tilde{r}_{\rm E}$,
where $\tilde{r}_{\rm E}$ is the 
radius of the Einstein ring projected into the observer's plane) and/or the
rotation of the binary lens 
assuming circular {face on} orbits (with an angular velocity of
$\dot{\beta}$ in units of {\rm deg~yr$^{-1}$}). 

The first stage of searching for the best-fit models was, however, done on
a grid more appropriate for uncovering repeating events. 
The grid covers a wide range for 6 parameters:  the
mass ratio, binary separation, minimum distances to the source trajectory
from the first and second mass and their corresponding times of minimum
approach. 
The starting values for the approach times are estimated from the
visual inspection of the light curve. For fixed distances from the two
masses, two source trajectories are possible depending on whether the
source trajectory intercepts the binary lens axis or not. Each parameter is
sampled with about 15-20 intervals. These initial searching parameters are
then transformed to  the ``standard'' model parameters mentioned above and
the $\chi^2$ is calculated. 
A few hundred models with low $\chi^2$'s from the grid are taken as initial
guess parameters and fed into a minimisation procedure based on the
Powell's method (\citealt{press92}).  

We also fit a simple static binary source model to each event. In this
model, the light curve is a sum of two standard single microlensing events
with two impact parameters  ($b_1$ and $b_2$), two parameters for the times
of maximum magnification ($t_{01}$ and  $t_{02}$), fractions of light
contributed to the total light by the two sources ($f_1$, and $f_2$,
$f_1+f_2 \le 1$), baseline magnitude ($I_0$) and Einstein radius crossing
time ($t_E$). To ensure the resulting models are comparable with the binary
lens models   a similar minimisation strategy is used with the same grid
sizes and  optimisation routines as those for the binary lens model. Notice
that the binary source model was fit only to non-caustic crossing events
since the binary source models cannot reproduce the sharp gradient features
in caustic-crossing events.  In addition, we apply an approximate wide
binary model following the concept  of \citet{1996ApJ...457...93D}. In this
model, the binary lens acts as two independent single lenses. If the two
individual magnifications are given as $\mu_1$ and $\mu_2$, then the
resulting magnification is approximated as  $\mu \approx \mu_1 + \mu_2
-1$. The fitting is done on a grid similar to that for the full binary lens
model. In practice it is important to fit both bumps simultaneously to
ensure the same constraints on the blending (and other) parameters.  
Due to the degeneracies in the blending model (\citealt{WP97}),
fitting each peak independently gives uncertain estimation of fluxes 
and timescales.

{Notice that the two (smooth) peaks produced by a wide binary lens
  correspond to the source approaches to the diamond shaped caustics, not
  to the lens components. Since both caustics lie between
the masses, fitting the simplified model systematically under-estimates
the binary separation and may give incorrect trajectory direction.}
However, since we are primarily
interested in the mass ratio of the binary which are given by the ratio of the
squares of the timescales of two magnification peaks
(\citealt{1996ApJ...457...93D}), this is not a significant limitation
(see \S\ref{results} and Fig. \ref{fig:ratios}).

\section{Results}
\label{results}
\label{sec:results}

\begin{table}
\caption{Candidates for repeating microlensing events.}
\begin{tabular}{lrrrl}
\hline
      Event OGLE- &$(f_1/f_2)_{\rm bs}$ & $q_{\rm bl}$ &  $q_{\rm approx}$ & Type  \\
\hline
      1999-BUL-42  & \multicolumn{3}{c}{model not found} & cc bl?       \\  %
      1999-BUL-45  & 0.189 & 0.203 & 0.270 & bl/bs   \\  
      2000-BUL-42  & 0.260 & 0.339 & 0.400 & bl      \\  
     2002-BLG-018  & 0.280 & 0.395 & 0.443 & bs/bl   \\  
     2002-BLG-045  & 0.045 & 0.008 & 0.011 & bl      \\  
     2002-BLG-128  &  --   & 0.611 & --    & cc bl   \\  %
     2003-BLG-063  & 0.539 & 0.203 & 0.428 & bl      \\  
     2003-BLG-067$^{\dag}$  & 0.817 & 0.788 & 0.844 & bs/bl   \\  
     2003-BLG-126  & 0.292 & 0.604 & 0.918 & bl      \\  
     2003-BLG-291  &  --   & 0.617 & --    & cc bl   \\  %
     2003-BLG-297  & 0.075 & 0.147 & 0.121 & bl/bs   \\  
     2004-BLG-075  & 0.707 & 0.587 & 0.664 & bl/bs   \\  
     2004-BLG-328  & 0.150 & 0.056 & 0.054 & bs      \\  
     2004-BLG-440  & 0.927 & 0.622 & 0.906 & bl      \\  
     2004-BLG-591  & 0.548 & 0.507 & 0.431 & bl      \\  
     2006-BLG-038  &  --   & 0.569 & --    & cc bl   \\  %
     2006-BLG-460  & 0.388 & 0.267 & 0.285 & bl      \\  
175257.97-300626.3 &  --   & 0.195 & --    & cc bl   \\  %
174828.55-221639.9 &  --   & 0.158 & --    & cc bl   \\  %
\hline
\end{tabular}

Light ratio from binary source model and mass ratios from full ($q_{\rm
bl}$) and approximate ($q_{\rm approx}$) binary lens models are shown. 
The last column indicates whether the binary lens (``bl'') or binary
source (``bs'') model is better. ``bl/bs" stands for comparable binary
lens and binary source models. Events with clear 
caustic crossing features (``cc bl") admit only the full binary lens model.
\label{table:events}
$\dag$ Modelling for this event indicates it did not fully return to the
baseline between the peaks (see the light curve in the
Appendix). Nevertheless, we include this event because of the lack of
sufficient data points around the ``saddle point'' between the peaks.
\end{table}

Table \ref{table:events} lists all 19 candidates for repeating events.
It shows also the mass ratios in the full and approximate wide binary
lens models and the light ratio from the binary source model. 
The last column in the Table indicates the nature of each event as
concluded from a comparison of the best $\chi^2$ values of the binary lens and binary source fits. 
The two models are regarded as comparable when their $\chi^2$ values differ
by less than $10$.  
In our terminology, ``bl" stands for binary lens, ``bs" for binary source
and ``cc bl" indicates that  
the caustic crossing features are clearly visible in the light curve for
which, therefore, only the full binary lens model is fit. 
The parameters of the best-fit binary lens and binary source models are
listed respectively in Tables \ref{tblmodels} and \ref{tbsmodels}.
The light curves of all the events are shown in the Appendix.

{The smooth (non-caustic-crossing) light curves with two peaks quite
  often have concurrent, similar quality fits based on binary source and
  binary lens models. This degeneracy is also known for events with much
  shorter duration, when the peaks partially overlap 
  (\citealt{Jar04, Jar06, 2007AcA....57..281S,Coll04}). Since binary
  lenses offer a much larger variety of possible light curve shapes,
  their fits are usually formally better.}

Clear caustic crossing features can be seen in the light curves of about
30 per cent of our candidates. This fraction is an order of magnitude
higher than that 
in the whole sample of microlensing events -- in the uniform sample of
3159 candidates from the EWS from OGLE-III, 73 events ($\sim
2.4\%$) show clear caustic crossing features in their light curves. 
Our simulations (section \ref{eff}) show that assuming all our candidate
events were 
caused by wide binary lenses we should expect that approximately
10-30 per cent should exhibit caustic crossing features (the fraction
increases as the mass ratio decreases). 
The high fraction of caustic crossing events in our sample of repeating
events suggests that the binary lens scenario is favoured for the majority
of our candidates. 

{The number of binary source events in our sample appears to be too low when
  compared with predictions of \cite{Griest92}, \cite{Han98},
  \cite{Dominik98}. Even if all ambiguous cases were classified as binary
  source events, we would get probability of only 0.15\% for their
  occurrence, much lower than the percentage from these studies. Our
  sample, however, includes only the events with well separated peaks,
  while the other authors include close binary sources, which have the
  highest probability of producing an event distinguishable from ordinary
  microlensing light curve.}

\begin{table*}
\begin{minipage}{142mm}
\caption{Parameters of the best-fit binary lens models.}
\begin{tabular}{lrrrrrrrrr}
\hline
       Event               & $\chi^2$/DOF & $q$  &  $d$  & $\beta$ &  $b$  &  $t_0$ & $t_E$ & $f$  & $I_0$ \\
\hline
       1999-BUL-42         & \multicolumn{9}{c}{=2003-BLG-220, model not found, see text}       \\  %
       1999-BUL-45         &   645.4/405 & 0.203 & 4.104 &   18.00 &  0.47 & 1401.6 &  30.2 & 1.00 & 17.75 \\  
       2000-BUL-42         &  8852.0/697 & 0.339 & 5.308 &   22.40 &  1.09 & 2096.7 &  99.0 & 0.37 & 13.58 \\  
      2002-BLG-018         &   142.9/111 & 0.395 & 6.572 &   -1.30 &  0.44 & 2510.3 &  34.6 & 0.91 & 18.06 \\  
      2002-BLG-045         &  3222.7/633 & 0.008 & 3.958 &  183.49 & -0.22 & 2360.8 &  26.4 & 0.97 & 18.76 \\  
      2002-BLG-128         &  2218.6/810 & 0.611 & 1.908 &  173.84 &  0.22 & 2463.2 &  59.5 & 0.12 & 17.74 \\  %
      2003-BLG-063         &  1105.4/317 & 0.203 & 3.154 &   25.76 &  1.39 & 2869.3 &  54.5 & 0.78 & 16.13 \\  
      2003-BLG-067         &  1236.4/323 & 0.788 & 3.636 &    0.60 &  0.47 & 2942.4 &  88.6 & 0.62 & 16.45 \\  
      2003-BLG-126         &  1027.9/295 & 0.604 & 2.808 &  -34.80 & -0.92 & 2804.2 &  22.8 & 1.00 & 15.66 \\  
2003-BLG-291$^\dag$        &   912.8/251 & 0.617 & 3.041 &  184.70 &  0.50 & 2925.8 &  43.5 & 0.38 & 17.45 \\  
      2003-BLG-297         &   984.6/297 & 0.147 & 2.912 &  179.33 &  0.51 & 2963.3 &  75.5 & 1.00 & 17.31 \\  
      2004-BLG-075         &   555.5/203 & 0.587 & 4.456 &  -11.09 & -0.19 & 3097.8 &   9.7 & 1.00 & 18.82 \\  
      2004-BLG-328         &  2782.3/545 & 0.056 & 3.399 &  175.70 &  0.20 & 3181.5 &  25.3 & 0.20 & 18.45 \\  
      2004-BLG-440         &  2220.0/527 & 0.622 & 4.499 &  158.66 &  0.65 & 3183.1 &   8.6 & 0.36 & 16.34 \\  
      2004-BLG-591         &  2478.3/672 & 0.507 & 2.874 &    4.69 &  0.26 & 3426.9 &  81.4 & 0.29 & 18.48 \\  
      2006-BLG-038         &  3556.7/744 & 0.569 & 3.138 &  206.47 & -0.68 & 3809.4 &  12.7 & 1.00 & 16.44 \\  %
      2006-BLG-460         &  2418.9/862 & 0.267 & 3.354 &  178.85 &  0.11 & 3982.7 &  20.0 & 0.69 & 19.10 \\  
175257.97-300626.3$^\ddag$ &  4845.1/932 & 0.195 & 3.268 & -171.58 & -0.37 & 2547.4 & 102.9 & 0.35 & 18.19 \\  %
174828.55-221639.9$^\S$    &   406.2/192 & 0.158 & 2.431 &  172.48 & -0.06 & 2827.7 & 155.2 & 0.31 & 18.93 \\  %
\hline
\end{tabular}
The columns are respectively, the name of the OGLE event,
$\chi^2$/number of degree of freedom (DOF), mass ratio $q$, binary separation $d$ (in units of the Einstein radius),
direction of the source trajectory with respect to the binary axis $\beta$ (in degrees),
impact parameter $b$, time of the closest approach to the center of mass $t_0$ (Julian date shifted by 2450000),
Einstein radius crossing time $t_E$ (in days), blending parameter $f=F_s/F_0$ and I-band brightness of the baseline $I_0$.\\
$\dag$ The model for the event 2003-BLG-297 is taken from Jaroszy{\'n}ski et al. (2005) which includes binary rotation.
$\ddag$ model with parallax motion, $\pi_E=0.279$.
$\S$ model with binary axis rotation, $\dot{\beta}=0.032^\circ/{\rm day}$.
\label{tblmodels}
\end{minipage}
\end{table*}

\begin{table*}
\begin{minipage}{142mm}
\caption{Parameters of the best-fit binary source models.} 
\begin{tabular}{lrrrrrrrrr}
\hline
            Event       & $\chi^2$/DOF &$b_1$ & $b_2$ &$t_{01}$ &$t_{02}$ & $t_E$ & $f_1$ & $f_2$ & $I_0$ \\
\hline
            1999-BUL-42 & \multicolumn{9}{c}{---}                                                         \\  %
            1999-BUL-45 &   646.4/405 & 0.450 & 0.546 & 1310.64 & 1420.33 &  35.8 & 0.109 & 0.576 & 17.75 \\  
            2000-BUL-42 &  8922.8/697 & 0.539 & 2.386 & 1748.74 & 2221.12 &  66.5 & 0.207 & 0.793 & 13.58 \\  
           2002-BLG-018 &   142.2/111 & 0.473 & 0.428 & 2350.71 & 2573.40 &  31.9 & 0.218 & 0.778 & 18.06 \\  
           2002-BLG-045 &  3269.6/633 & 0.220 & 0.002 & 2359.97 & 2457.28 &  25.8 & 0.957 & 0.043 & 18.76 \\  
           2002-BLG-128 & \multicolumn{9}{c}{---}                                                         \\  %
           2003-BLG-063 &  1198.6/317 & 0.586 & 1.675 & 2748.84 & 2891.19 &  29.2 & 0.350 & 0.650 & 16.13 \\  
           2003-BLG-067 &  1228.8/323 & 0.420 & 0.451 & 2771.86 & 3077.76 &  83.9 & 0.273 & 0.334 & 16.45 \\  
           2003-BLG-126 &  1221.7/295 & 0.136 & 1.120 & 2774.55 & 2831.55 &  15.4 & 0.774 & 0.226 & 15.66 \\  
           2003-BLG-291 & \multicolumn{9}{c}{---}                                                         \\  %
           2003-BLG-297 &   989.4/297 & 0.437 & 0.295 & 2934.13 & 3143.78 &  78.6 & 0.682 & 0.051 & 17.31 \\  
           2004-BLG-075 &   560.8/203 & 0.304 & 0.448 & 3072.38 & 3112.70 &   9.6 & 0.414 & 0.586 & 18.82 \\  
           2004-BLG-328 &  2710.3/545 & 0.240 & 0.004 & 3177.38 & 3255.48 &  26.1 & 0.204 & 0.031 & 18.45 \\  
           2004-BLG-440 &  2237.1/527 & 1.897 & 0.624 & 3169.26 & 3204.09 &   5.3 & 0.515 & 0.477 & 16.34 \\  
           2004-BLG-591 &  2489.5/672 & 0.046 & 0.200 & 3289.05 & 3499.46 & 106.8 & 0.063 & 0.115 & 18.48 \\  
           2006-BLG-038 & \multicolumn{9}{c}{---}                                                         \\  %
           2006-BLG-460 &  2455.2/862 & 0.099 & 0.054 & 3969.47 & 4030.86 &  22.9 & 0.470 & 0.183 & 19.10 \\  
     175257.97-300626.3 & \multicolumn{9}{c}{---}                                                         \\  %
     174828.55-221639.9 & \multicolumn{9}{c}{---}                                                         \\  %
\hline
\end{tabular}
The columns show the name of the OGLE event, $\chi^2$/the number of DOF, impact parameters $b_1$ and $b_2$ for two source stars,
times of closest approach to both components $t_{01}$ and $t_{02}$, Einstein radius crossing time $t_E$,
blending parameters $f_1=F_{s1}/(F_{s1}+F_{s2}+F_b)$ and $f_2=F_{s2}/(F_{s1}+F_{s2}+F_b)$, and I-band brightness  of the baseline $I_0$.
$F_{s1}$ and $F_{s2}$ are the fluxes from the two sources, and $F_b$ is the flux from any other unrelated star(s) within the seeing 
disc (blend).
\label{tbsmodels}
\end{minipage}
\end{table*}

\begin{figure}
\begin{center}
\includegraphics[width=\colwidth]{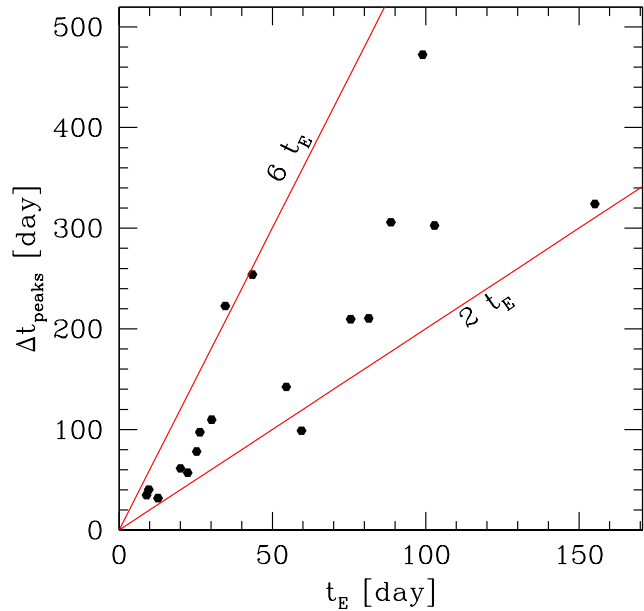}
\caption{Time between two peaks in repeating events versus the Einstein radius crossing time, $t_E$.}
\label{fig:times}
\end{center}
\end{figure}

One practical issue one needs to know for future search for repeating
events is how long an observer has to wait for the secondary episode to occur.
This is illustrated in Figure \ref{fig:times} showing the time between the two peaks vs. the event timescale. 
As predicted in \citet{1996ApJ...457...93D} this time is of the order of a few (from 2 to 6) Einstein-radius crossing times. 
For the repeating events described here this is between 32 to 472 days with a median of 142 days.

\begin{figure}
\begin{center}
\includegraphics[width=\colwidth]{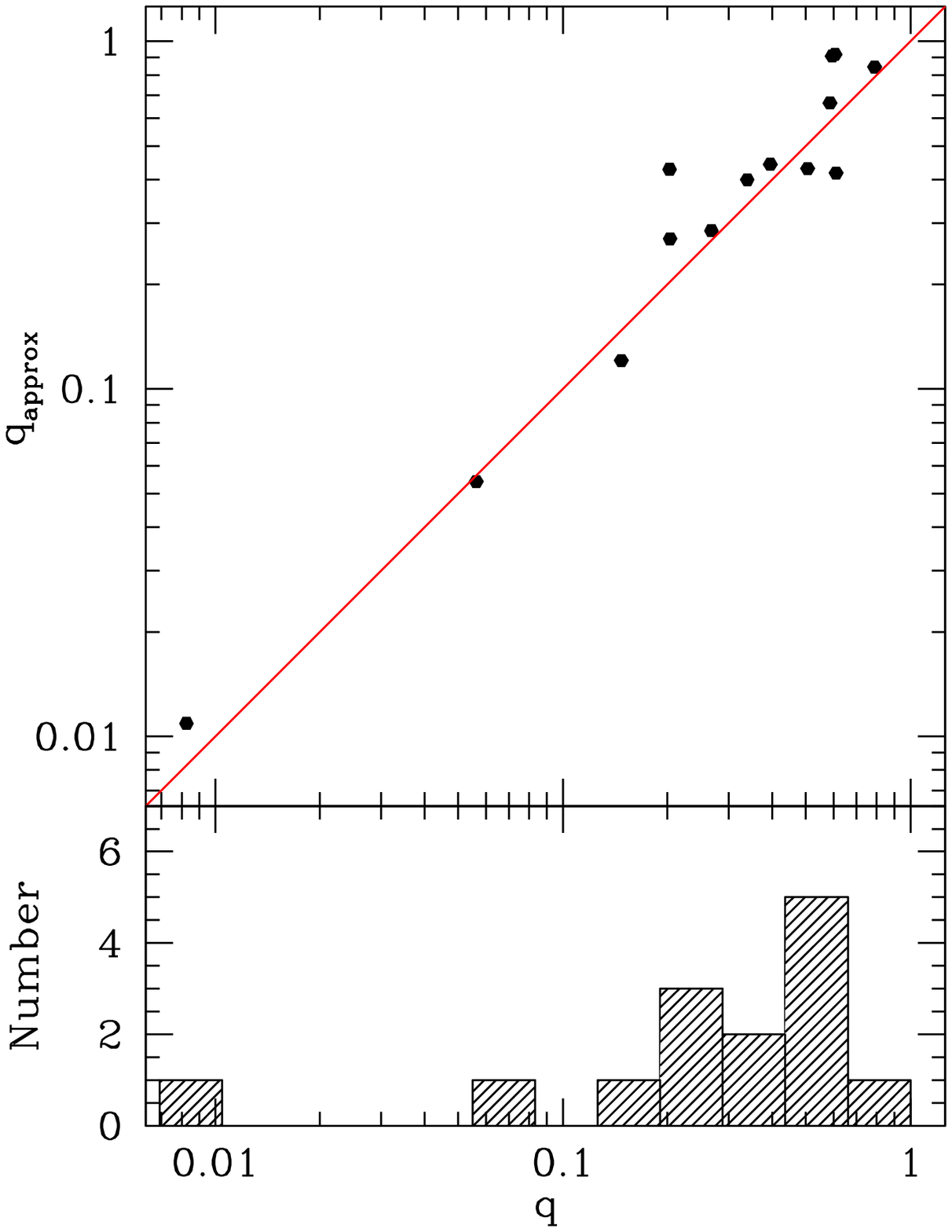}
\caption{Correlation between the mass ratios obtained in the full ($q$)
and approximate ($q_{\rm approx}$) binary lens models (upper panel) and
the observed histogram of mass ratios derived in the full binary lens model (lower panel).}
\label{fig:ratios}
\end{center}
\end{figure}

Since the approximate binary model gives a quick way to estimate the binary mass ratios (see \S\ref{sec:model}), it is important to check its accuracy. 
Figure \ref{fig:ratios} compares the mass ratios in the full and
approximate binary lens models where both fits are available. 
A strong correlation is clearly visible (with an RMS difference of about
36\%), indicating that the simplified model can indeed be used to
estimate the mass ratio in repeating events due to wide binary lenses.

\subsection{Distribution of binary lens mass ratios} 
\label{eff}

The histogram in the bottom panel of Fig. \ref{fig:ratios} shows the
distribution of the mass ratio in the wide binary lens models. It shows a
peak around $q \sim 0.5$ and a decrease for small values of $q$. The
decrease is, however, likely due to a larger incompleteness at small $q$ in
the observations. Thus to uncover the true distribution of mass ratios, we
have to correct the histogram in Fig. \ref{fig:ratios} for the detection
efficiency. 

\subsubsection{Simulations of detection efficiency for wide binary lenses}

A rigorous detection efficiency analysis at the pixel level involves
many factors and is beyond the scope of the paper. However, since our
sample is small and Poisson errors are large (see Fig. \ref{fig:eff}), a
first-order approach at the catalog level is sufficient to 
account for the primary selection effects for wide binaries with different mass ratios.

We perform simulations of repeating events due to wide-separation
lenses with the mass ratio ranging from $10^{-3}$ to 1 at 25 uniform,
logarithmically-spaced intervals. Depending on the
mass ratio, about $10^4-10^5$ binary lensing events are generated.
An automated algorithm for identifying the fraction of repeating events observed in the sample is then used (see \S\ref{sec:eff}).

To create a synthetic microlensing event, we first 
generate a physical model of a binary lens and source trajectory and then
based on the 
physical model, a light curve with sampling rates similar to those of
OGLE is simulated. To generate a binary microlensing model the
following steps are performed:
\begin{enumerate}
\item For a given mass ratio, a binary separation is drawn 
from the range 1 to 36 Einstein radii using a uniform distribution in
logarithm. 
{Our simulations show that the detection efficiency falls to zero
outside this range. For lenses of masses $M \in [0.1, 1]~M_\odot$ placed in
the Galactic disk and sources in the bulge, the above range corresponds
roughly to physical separations of $1$ -- $100~\mathrm{AU}$.}
\item To ensure at least one magnification episode, we set the source
trajectory to pass the heavier lens at an impact parameter from 0 to 1
Einstein radius. 
\item The trajectory angle is set randomly from 0 to 360 degrees.
\item The time of the closest approach to the heavier mass is chosen
from a uniform distribution to lie within the whole observing time.
\item The Einstein-radius crossing time is drawn from an approximate
distribution that matches that of the observed OGLE lenses: 
a log-normal distribution centered on $t_E=26$ days with a standard
deviation of 0.3. 
\end{enumerate}
After the physical model is chosen using the procedure described above, the
magnification as a function of time can be easily found. However, to mimic
an observed light curve, several further steps are necessary.
\begin{enumerate}
\item The baseline magnitude is chosen randomly from the luminosity
  function of one of the OGLE fields (BLG104.6). 
\item {The value of the blending parameter is chosen at
  random between 0 and unity. The small sample of binary lens models (\citealt{Jar04}) supports
 such a choice of the blending parameter distribution.   The values
  of the baseline flux and blending parameter combined with the theoretical
  lens model yield a perfect light curve without errors.} 
\item To account for observational gaps in the light curve,
the epochs of observations are chosen so as to match
either BUL\_SC34.I.36546 in OGLE-II or BLG104.6.I.7723 in OGLE-III. They
have been chosen as a reference for epochs and observational errors (to
be used in eq. \ref{eq:deltaI}) as  their light curves span from 1998 to 2007 during which all our
repeating events candidates have been found.
\item Each measurement has an observational error $\Delta I$ 
from the rescaled values for a reference star $\Delta I_{\rm ref}$ 
using the empirical formula from \cite{PhD}:
\begin{equation}
\Delta I = \Delta \, I_{\rm ref} 10^{0.33875 ( I - I_{\rm ref} ) }
\label{eq:deltaI}
\end{equation}
where $I$ is the model magnitude, $I_{\rm ref}$ is the magnitude 
and $\Delta I_{\rm ref}$ is the error bar for the reference star at the epoch.
\item Gaussian errors are added to the light curve. In this step, we take
  into account the fact that OGLE errors are slightly 
underestimated: for every simulated data point, a Gaussian standard
  deviation is derived based  
on its error bar value $\Delta I$ in eq. (\ref{eq:deltaI}) using
\begin{equation}
\sigma = \sqrt{ ( 1.38 \Delta I )^2 + 0.0052^2 }.
\end{equation}
This rescaling has been obtained following the method of \citet{Wyrz2008}
by comparing 
the {\it rms} of a set of constant stars to
{their mean error bars}
returned by the photometry pipeline.
\end{enumerate}

\subsubsection{Efficiency-corrected statistics of binary mass ratios}
\label{sec:eff}

To calculate the detection efficiency of repeating events for a given mass
ratio, all the light curves of synthetic events must be classified.  
To accomplish this, we fit a standard Paczy{\'n}ski model to all
the light curves and compare the resulting
$\chi^2$ with a constant line fit. {We have also applied the same
procedure to check if a Paczy{\'n}ski model fit to a constant
luminosity light curve (with Gaussian errors) can improve the
fit. Our experiment shows that the typical improvement for $\sim 10^3$
d.o.f. is $\Delta\chi^2 \approx 15$ and exceeds 55 in $\sim 1\%$ of
cases. When investigating the synthetic light curves, we require
a $\Delta\chi^2>55$ improvement of the Paczy{\'n}ski model over the constant line
model to treat the event as a microlensing candidate. (Notice that while
each curve is simulated using microlensing, the gaps between the seasons and/or
sampling rate may prevent it from being ``discovered''.)}

\begin{figure}
\begin{center}
\includegraphics[width=\colwidth]{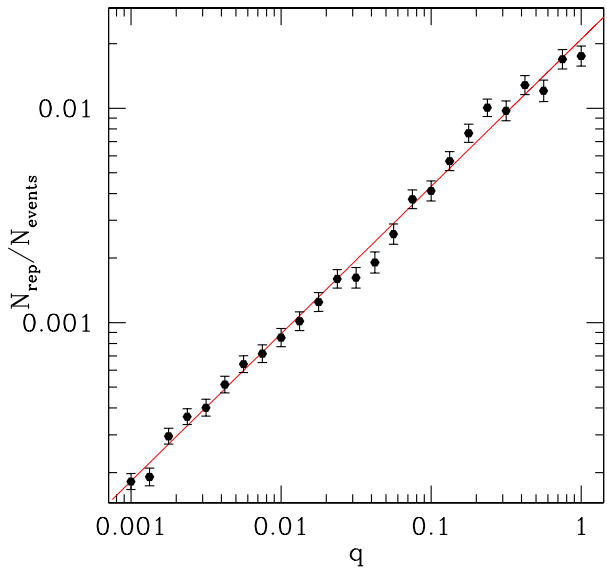}
\caption{Relative detection efficiency,  defined as the ratio of the
number of detected ``repeating'' events and the total number of detected microlensing events, as a function of the mass ratio $q$. 
Error-bars denote Poisson noise based on population of simulated
sample. {The solid line is an empirical fit of $N_\mathrm{rep}/N_\mathrm{event} \approx
0.021q^{0.687}$.}}
\label{fig:eff}
\end{center}
\end{figure}

We then use our automated algorithm (see \S\ref{automatic}) to identify
repeating events in our sample.  
The relative detection efficiency is then defined as the
ratio of repeating events ($N_{\rm rep}$) found in the sample of all
identified microlensing events ($N_{\rm events}$). Figure \ref{fig:eff}
shows the efficiency with respect to mass ratio ($q$). {It can be
approximated by the formula $N_\mathrm{rep}/N_\mathrm{event} \approx
0.021q^{0.687}$.}

\begin{figure}
\begin{center}
\includegraphics[width=\colwidth]{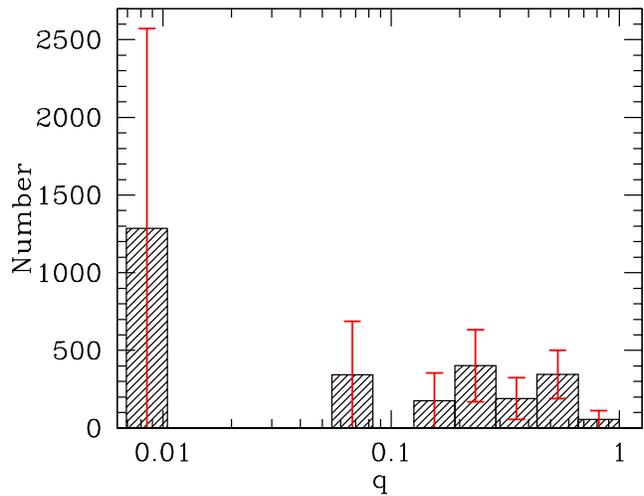}
\caption{Distribution of binary mass ratios after correcting for detection efficiency. 
Error bars are from the Poisson noise. The left-most bin at $q=0.01$ contains only one event. }
\label{fig:hist}
\end{center}
\end{figure}

{Fig. \ref{fig:hist} shows likely underlying distribution of binary mass
ratios obtained by convolution of observed mass ratios (Fig. \ref{fig:ratios}) 
and their detection efficiency (Fig. \ref{fig:eff}).
}
For mass ratios between $0.07 \la q \la 1$, the distribution is consistent
with a uniform distribution in the logarithm of mass ratio with Poisson 
errors, in agreement with earlier results of \cite{Trimble1990}. Although
we have one event (OGLE-2002-BLG-045) for the smallest mass ratio bin, 
because of the low detection efficiency for such events, the inferred 
relative number is quite high, but the error bar
is large. Nevertheless, it may indicate the presence of a
different population of binary (planetary) systems with extreme mass ratios.

Assuming a uniform distribution of binary lenses in $\log\,q$ and
$\log\,d $, we can calculate the averaged efficiency of the detection 
for  $q \in [0.1, 1]$ and $d \in [1, 36] r_E$, obtaining
$<N_\mathrm{rep}/N_\mathrm{events}> =0.0105$. Notice that we have
excluded the brown dwarf and planetary binary companions (with mass
ratios between $q \in [0.001, 0.1]$ since there is only one such
planetary candidate. If all the events were caused by binary lenses with parameters from this range, one would expect
$4120 \times 0.0105 \approx 43$ detections of repeating events.
Since twelve repeating events are unambiguously classified as binary lens cases, this suggests that at least 28\%
of all lenses are binary systems within the specified range in the mass
ratio and separation. Binary stars roughly follows a uniform
distribution per decade of separation
between $10^{11}$\,cm to $10^{17}$\,cm (\citealt{Abt83, MP91}), thus we expect a fraction of
$\log(36/1)/6=26\%$ within $d \in [1, 36] r_E$, in good agreement
with the 28\% we estimated above {\it if} all the lenses are in binary
systems. 
{However, earlier studies indicate that about 50\% of solar-type
stars are in binary stars, while for M stars this fraction may be even lower
(\citealt{DM91}; \citealt{Fisher92}; \citealt{RG97}; \citealt{Lada06}
and references therein), thus we would expect a factor of $\sim 2$
smaller number of repeating events due to binary lenses. There are
several possiblities to explain the 
``discrepancy''. 1) It may simply be due to small number Poisson fluctuation.
2) The binary fraction is under-estimated in spectroscopic surveys. 3)  
The distributions of the separation and mass ratio may differ
from being uniform per decade as we assumed (our data are
consistent with this assumption within the small number statistics.) 4) Some of the ``repeating'' events are actually not due to binary lenses.
In the future, with a much larger number of repeating events and denser
time samplings (e.g., with OGLE-IV), we will be in a better position to test
these different possibilities.}

\subsection{Comments on two individual events}

In this section, we study two particularly interesting ``repeating'' microlensing events.
OGLE-1999-BUL-42/OGLE-2003-BLG-220, and OGLE-2002-BLG-045 in more detail.

\subsubsection{OGLE-1999-BUL-42/OGLE-2003-BLG-220}

For this event, we were not able to find any 
satisfactory model. This interesting case requires closer scrutiny.
The light curve of the event in 1999 resembles a caustic crossing binary lens,
but because of the lack of sharp features it can also be explained by a binary 
source model. However, a static binary source model turns out to be insufficient.
The secondary bump from 2003 shows nearly no binarity features except one outlying data point and a slight asymmetry.

The timescales of both magnification episodes are of the order of 20 days
and the separation between them is very long, about 1400 days. This indicates it is unlikely that both bumps were produced by a 
very wide binary lens, because such lenses have tiny caustics and as
such they cannot explain the features in the light curve from 1999. 
One possible scenario for the repeatability is 
that the source was lensed by two unrelated lenses or two sources were lensed by one binary lens; the event may also be due to a variable star rather than microlensing.

We have attempted to gain some additional insights using information available from the OGLE photometry pipeline
using the Difference Image Analysis (DIA) \citep{wozniak}.
In this method all variable sources are detected on the subtracted image 
and for each variable the centroid of the subtraction residuals indicates the real position of the variable star. 
Because of blending this position may not be aligned with the position of the baseline star on the template image.
In the OGLE database (A. Udalski, private communication) we checked the positions of the residuals
with respect to the template positions around the 1999 and 2003 peaks.
Because of sparse sampling there are only a few high signal-to-noise measurements in the 2003 peak,
however we can still confirm that the position of the lensed source is consistent with the centre of the blend. 
This does not necessarily mean there was no blending, but simply that the source happens to be located near the 
centre of the blended object. On the other hand, the 1999 peak had a much better coverage and the detected
astrometric signal is firm, showing a clear displacement of the residuals in respect to the blended position by about 150-200 mas. 
This result indicates that it is unlikely the 1999 and 2003 events were caused by the same lens, unless it moves
very fast ($\sim 40\,{\rm mas/yr}$) and/or is a very close object.
The remaining (microlensing) hypothesis is that it is a pair of completely unrelated events occurring on two different sources.
Further studies, including high resolution imaging (e.g. with the Hubble
Space Telescope), would either confirm or reject this scenario.

\subsubsection{OGLE-2002-BLG-045} 

The second brightening of this event was much shorter than the first 
one and the binary lens model gives a low mass ratio of $q=0.008$ (the approximate binary lens model gives
$q=0.011$). For a typical lens mass of about $0.3 M_\odot$, this would
imply a planet mass of a few Jupiter masses.
Unfortunately, sparse sampling of both peaks prevent us from concluding convincingly the planetary nature of the lens.
Nevertheless this demonstrates repeating events may be a potentially
exciting channel of detecting planets in repeating events (e.g.,
\citealt{stefano99}).

\subsection{Contaminants to the OGLE Early Warning System}

As a by-product of our search, we identified a number of
non-microlensing events in the database. In total we found 64
mis-classified events out of 3159 events (about 2\%) in the
OGLE-III microlensing candidates identified by the EWS.
This low rate is re-assuring as it indicates that EWS, partially based
on human interpretation, is highly reliable in discovering true
microlensing events. Dwarf novae are the main contaminators (24). 
Thanks to the 15 years of continuous monitoring of OGLE, it is possible
to detect up to 30 outbursts for one dwarf nova. However, the majority of our stars had 2-4 observed outbursts. 
It is also possible to measure very long pseudo-periods of outbursts -
for 5 of our dwarf novae candidates the time between outbursts is longer than 1000 days.
Note that all dwarf novae found with the visual inspection are also retrieved by our automated fitting algorithm.
Thus it is possible to use this approach to identify these stars in the
entire OGLE photometric database; we plan to conduct such a study in the near future.

\section{Conclusions} 
\label{concl}

A small fraction ($\approx 0.5\%$) of microlensing events do repeat: our
search revealed 19 repeating candidates in the sample of 4120 microlensing events we studied.
This gives a rate of $\sim$ {3} events per year in the OGLE database at
the current discovery rate of $\sim$ 600 events/yr. Both the high fraction of caustic crossing events and comparison of the 
goodness of fit between binary lens and binary source models
suggest that probably most of these are due to wide binary lenses,
{although the predicted number seems to be somewhat higher than expected
(see \S\ref{sec:eff}).}

We find that it is possible to estimate the mass ratios for repeating events 
in a straightforward manner. With a growing number of microlensing events observed every 
year these events could provide a valuable sample to study
the stellar binary populations in the Galaxy. This method operates in
the time domain, different from the usual spectroscopic studies of binaries in the
solar neighbourhood (e.g., \citealt{Fisher05}). Accounting for the
detection efficiency, the distribution of mass ratios of binaries
appears to be consistent  with a uniform logarithmic distribution, in
agreement with that from previous spectroscopic studies (\citealt{Trimble1990}).

Our study also illustrates an example of a missed opportunity to find a planet (OGLE-2002-BLG-045). 
The second brightening episode was much shorter than the first. 
Unfortunately the sampling was not dense enough after the first episode returned to the baseline.
In the future it would be profitable for survey teams and follow-up networks to
pay more attention to microlensing events even after the main magnification peak (the median time between the two peaks is about 5 months).
However, given the limited observational resources, this is difficult to implement in
the current mode of extrasolar planet discovery with alerts from survey teams followed by other teams with intensive observations. 
Nevertheless, the number of repeating events will increase considerably
with next-generation wide-field microlensing surveys from the ground
(e.g. \citealt{Gould07}) and from space (\citealt{Bennett07}). The
ground-based experiments are already evolving toward a wide-field
network, starting with the upgraded MOA-II experiment and the soon-to-be upgraded OGLE project
(OGLE-IV). The dense sampling will be particularly important for the detection of extrasolar 
planets on wide orbits and will offer a new channel for extrasolar planets
discovery (\citealt{stefano99}). For these events 
the mass ratio can be approximately ``read" off from the light curve using the approximate binary lens model.
When the next-generation microlensing experiments become fully operational, this method may become fruitful.

\section*{Acknowledgments}

This work would not be possible without the help from Prof. Andrzej Udalski and the rest of the
OGLE Team who collected and reduced the 15 years of observational
data and allowed us to use their unpublished data.
We thank Drs. Martin Smith, Szymon Koz{\l}owski and Nicholas Rattenbury
for support and discussions, and {the referee Dr. Scott Gaudi for a thorough report that improved the paper.}
JS, {\L}W and SM acknowledge support from the European Community's FR6 Marie
Curie Research Training Network Programme, Contract No. MRTN-CT-2004-505183
``ANGLES". JS thanks Prof. Ian Browne and Dr. Wyn Evans for an
opportunity to visit the UK where this study was initiated.
This work was supported in part by Polish MNiSW grants N203 008 32/0709 and N203 030 32/4275.


\bsp

\appendix

\section[]{Light curves of 19 repeating candidate events together with the best-fit binary lens model.}

\begin{scriptsize}

\noindent
\includegraphics[width=\colwidth]{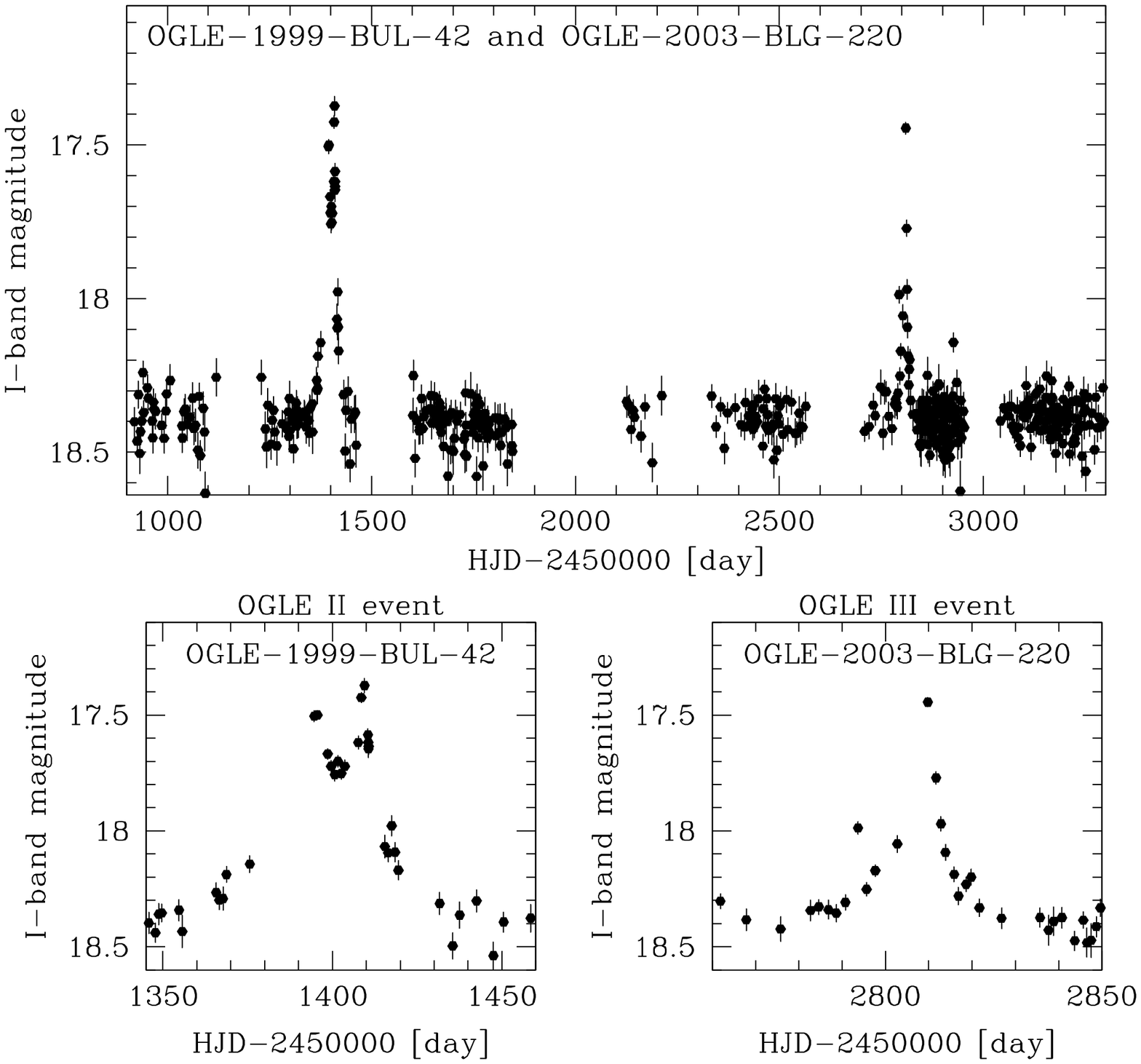}

\noindent
\includegraphics[width=\colwidth]{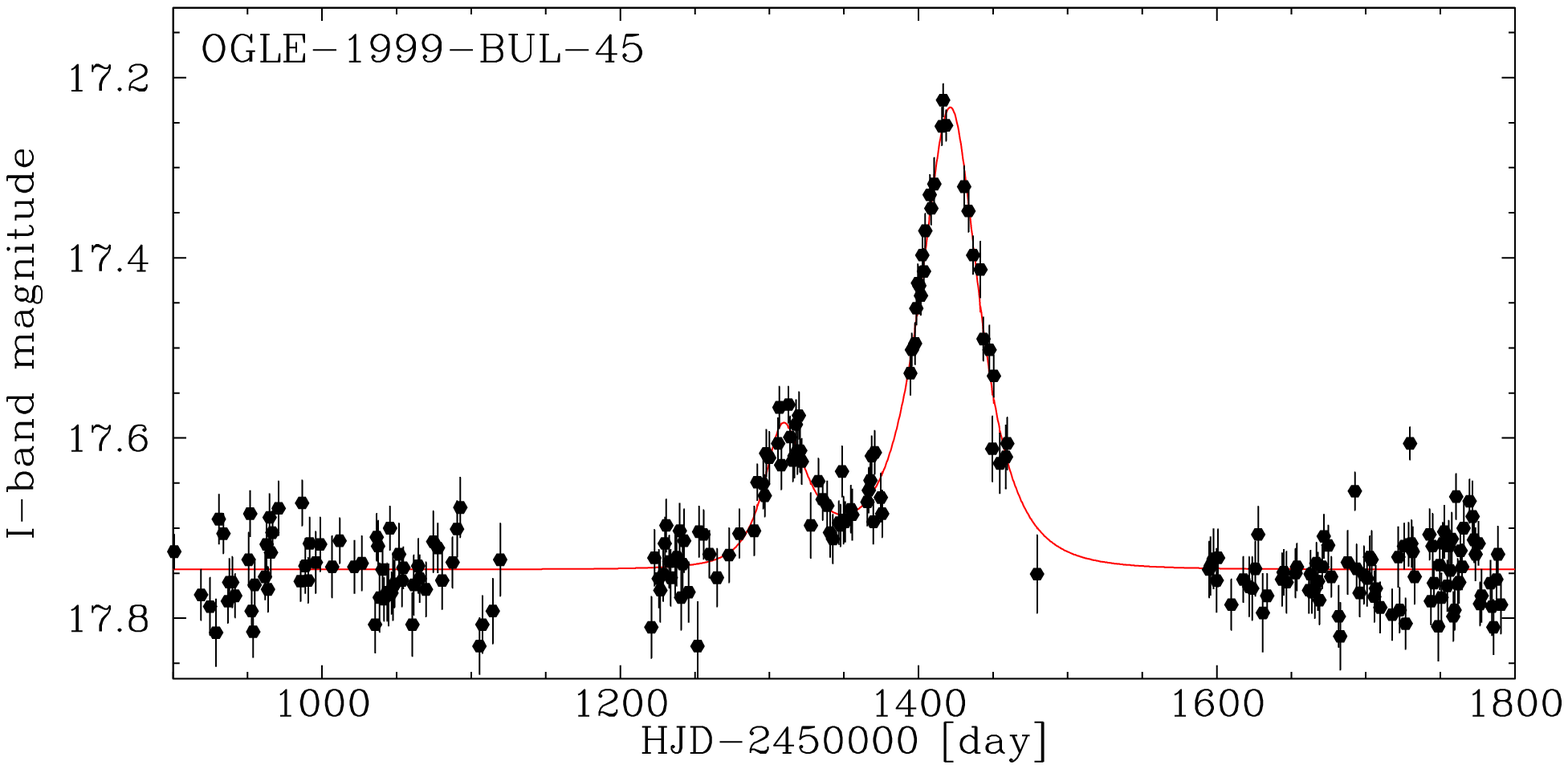}

\noindent
\includegraphics[width=\colwidth]{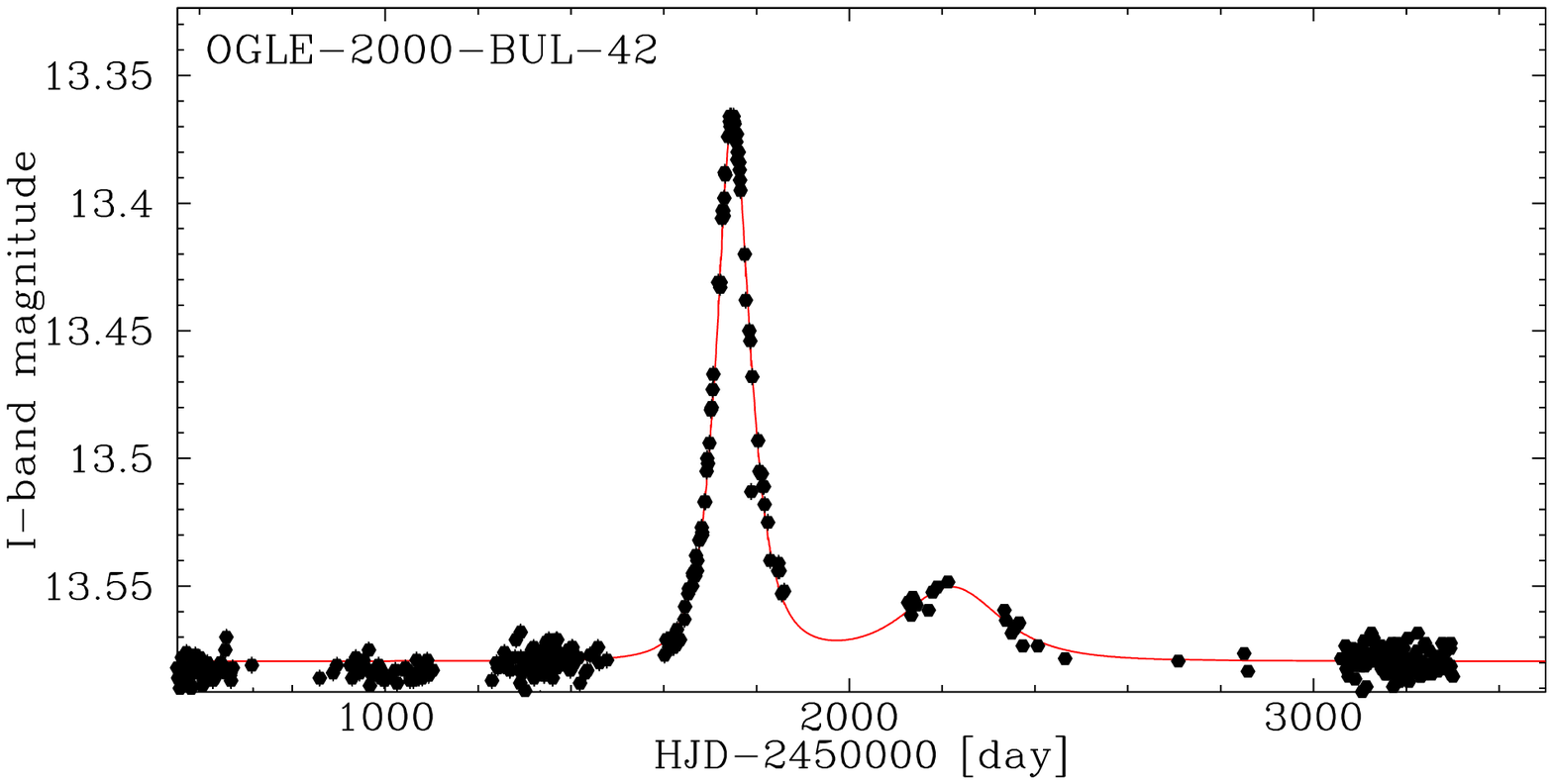}

\noindent
\includegraphics[width=\colwidth]{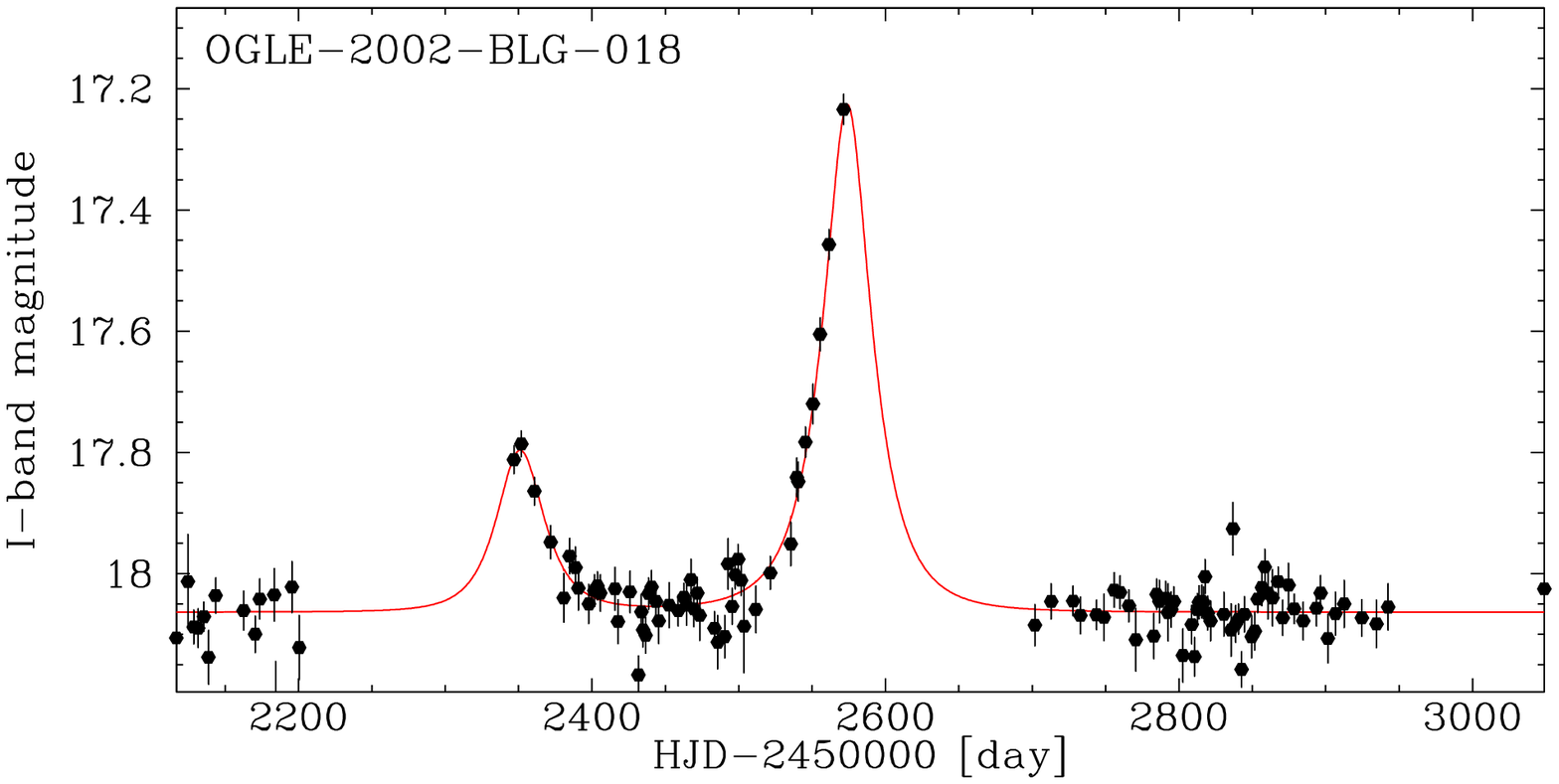}

\noindent
\includegraphics[width=\colwidth]{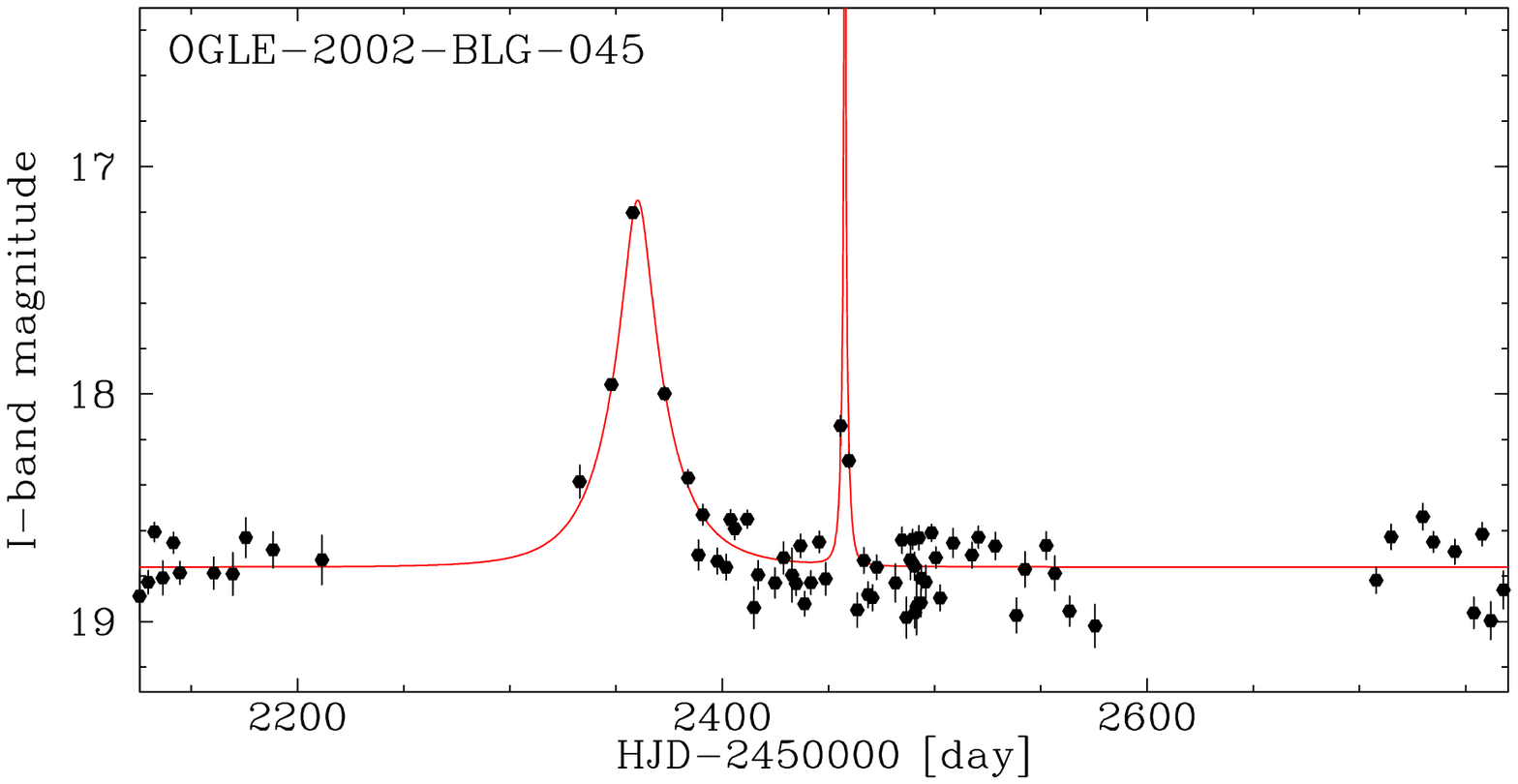}

\noindent
\includegraphics[width=\colwidth]{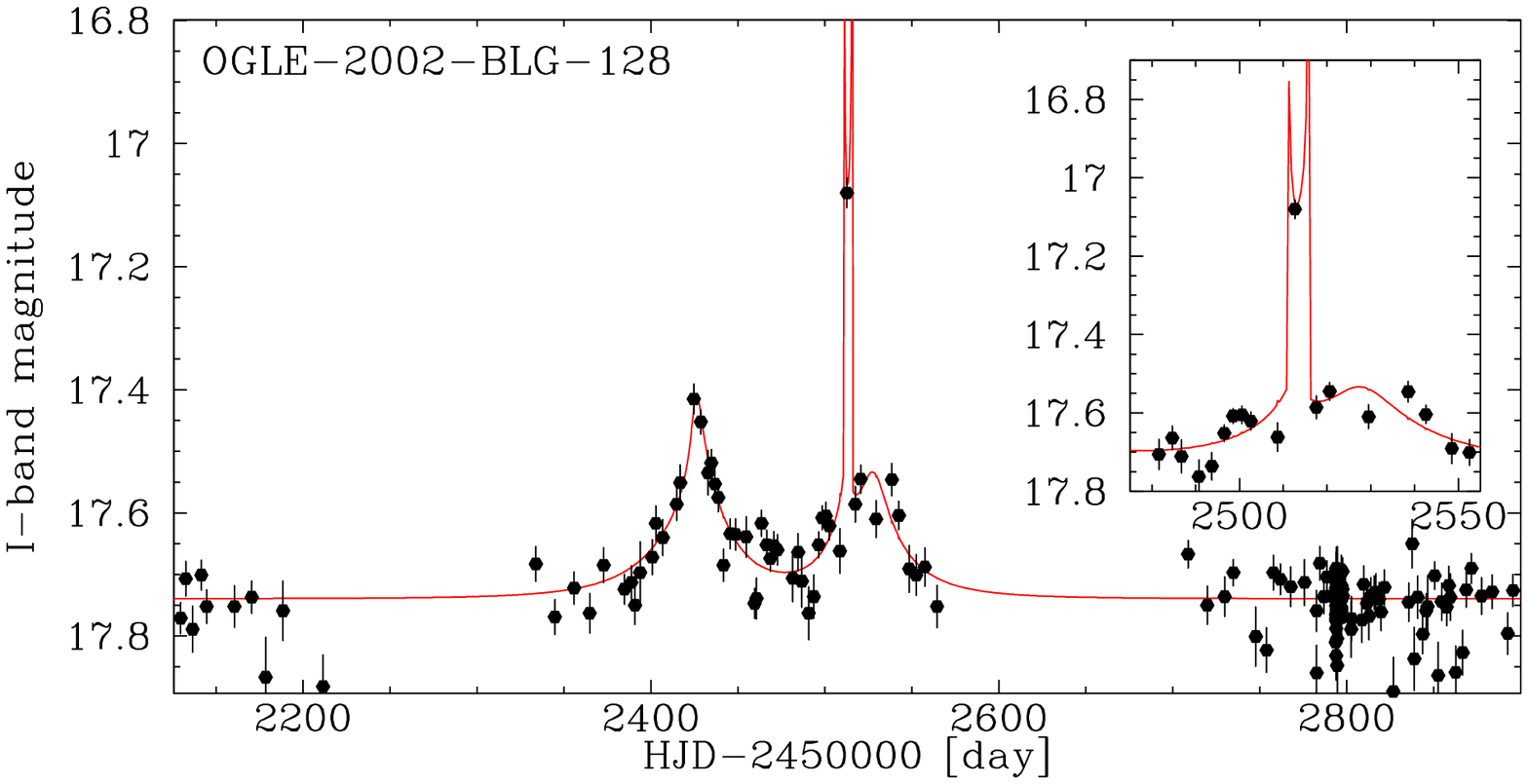}

\noindent
\includegraphics[width=\colwidth]{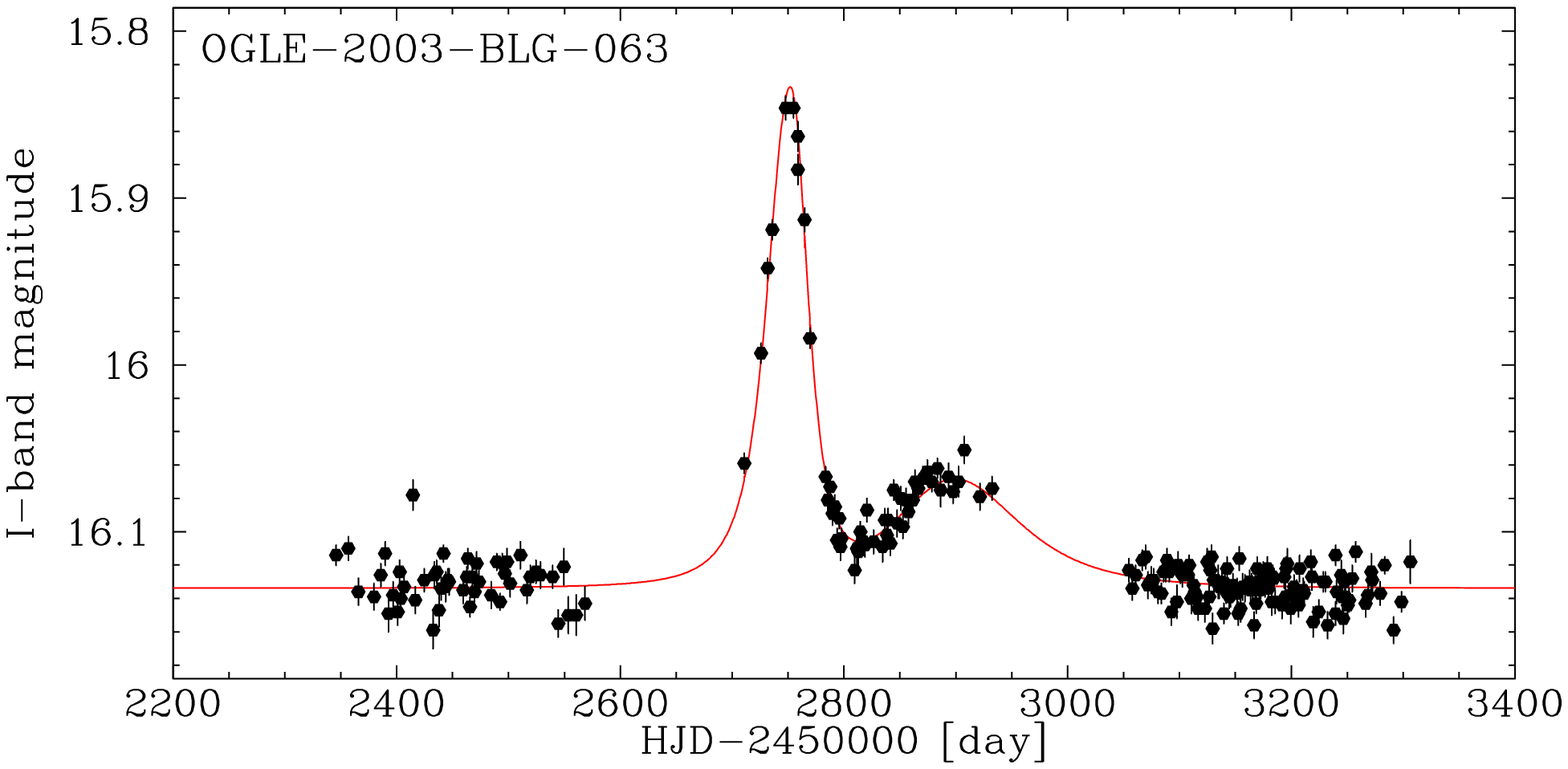}

\noindent
\includegraphics[width=\colwidth]{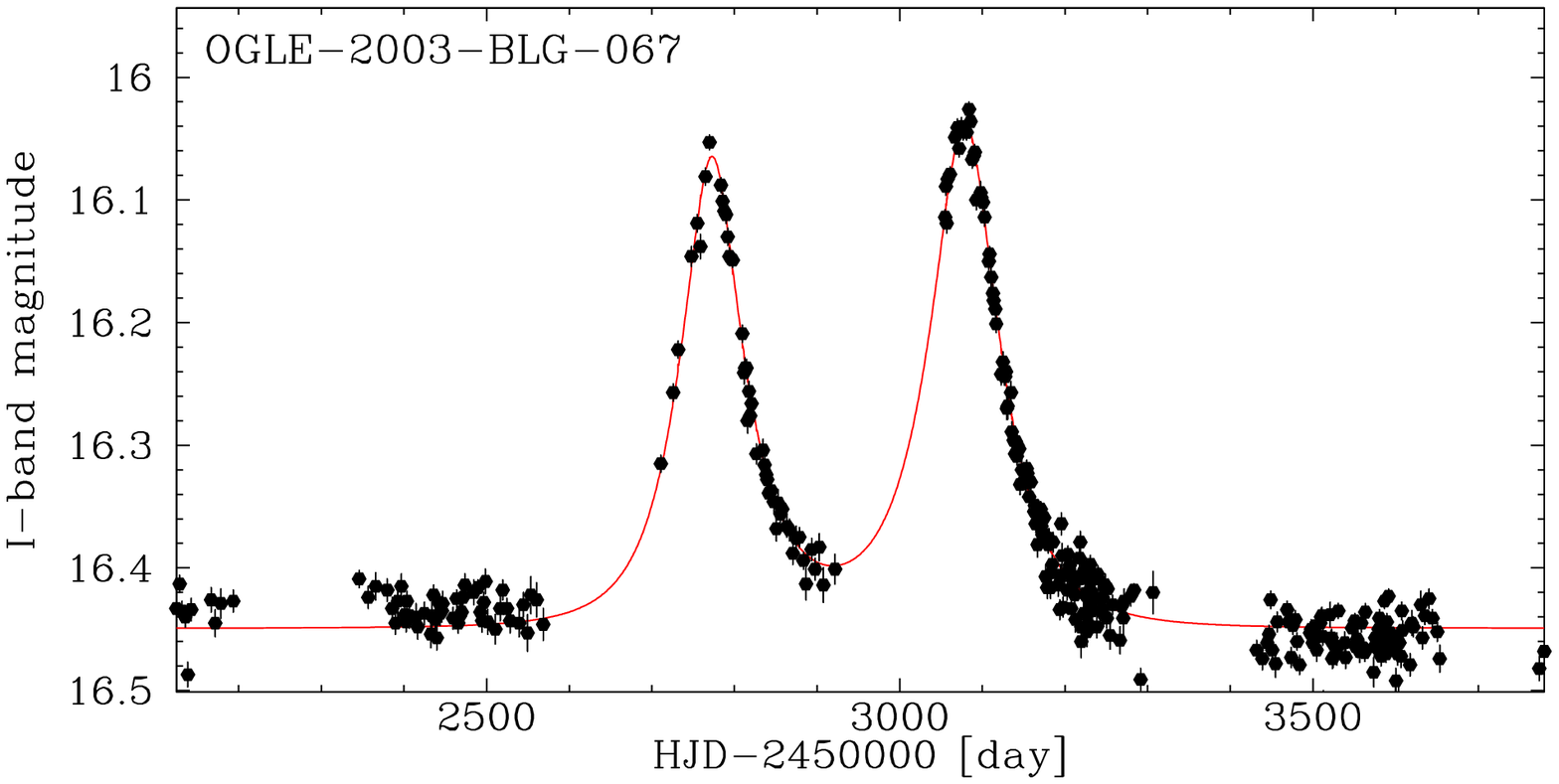}

\noindent
\includegraphics[width=\colwidth]{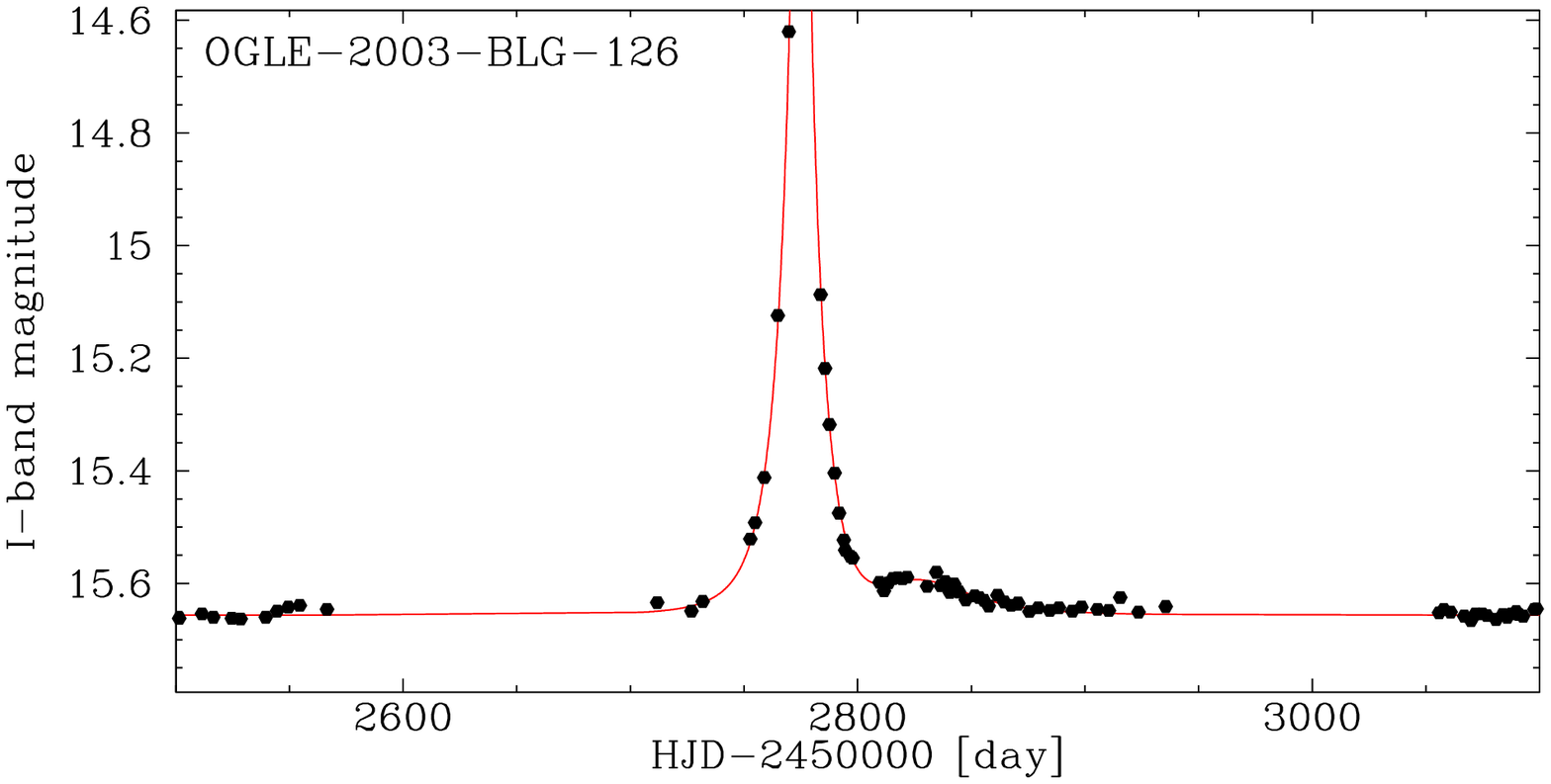}

\noindent
\includegraphics[width=\colwidth]{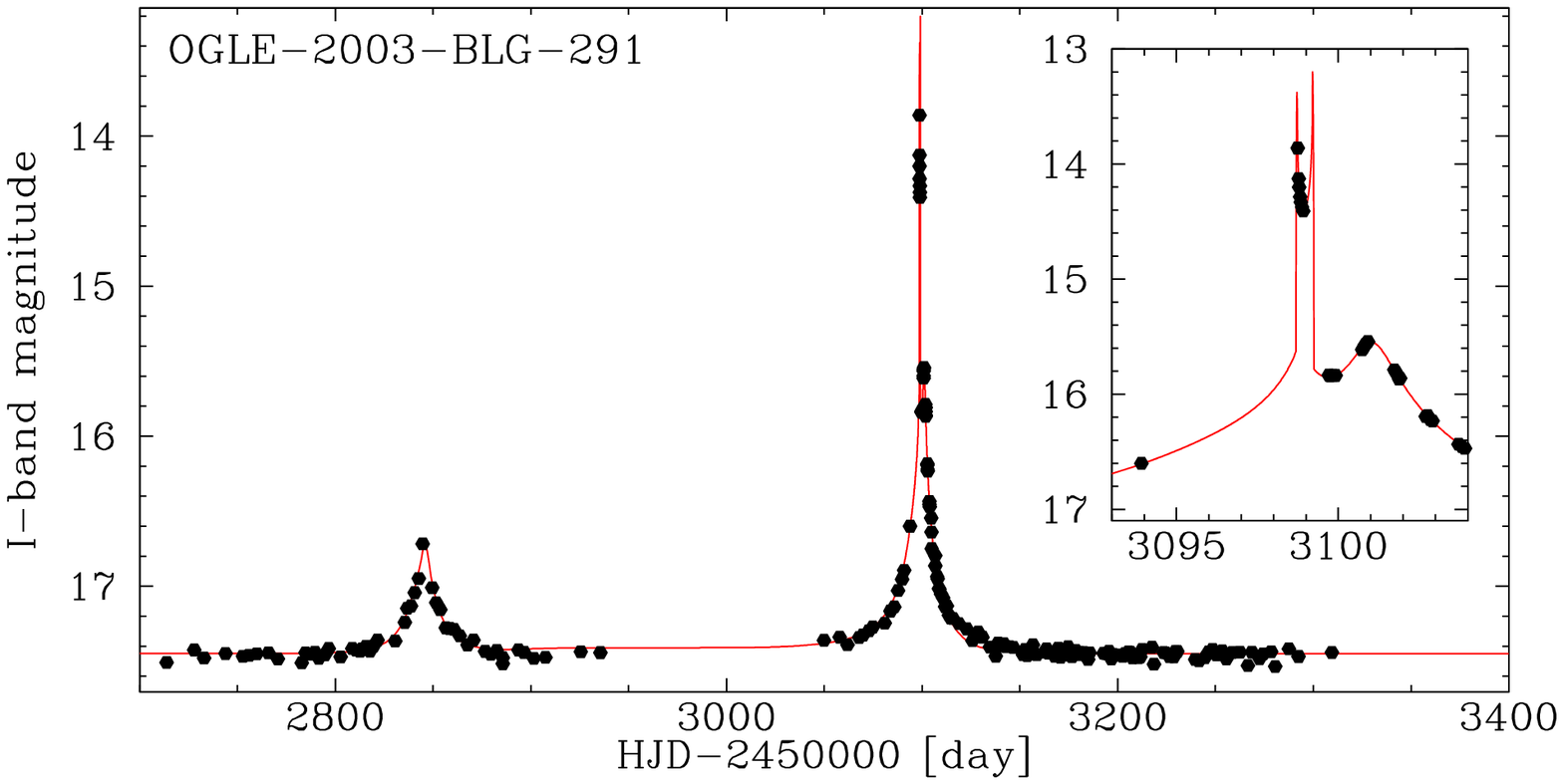}

\noindent
\includegraphics[width=\colwidth]{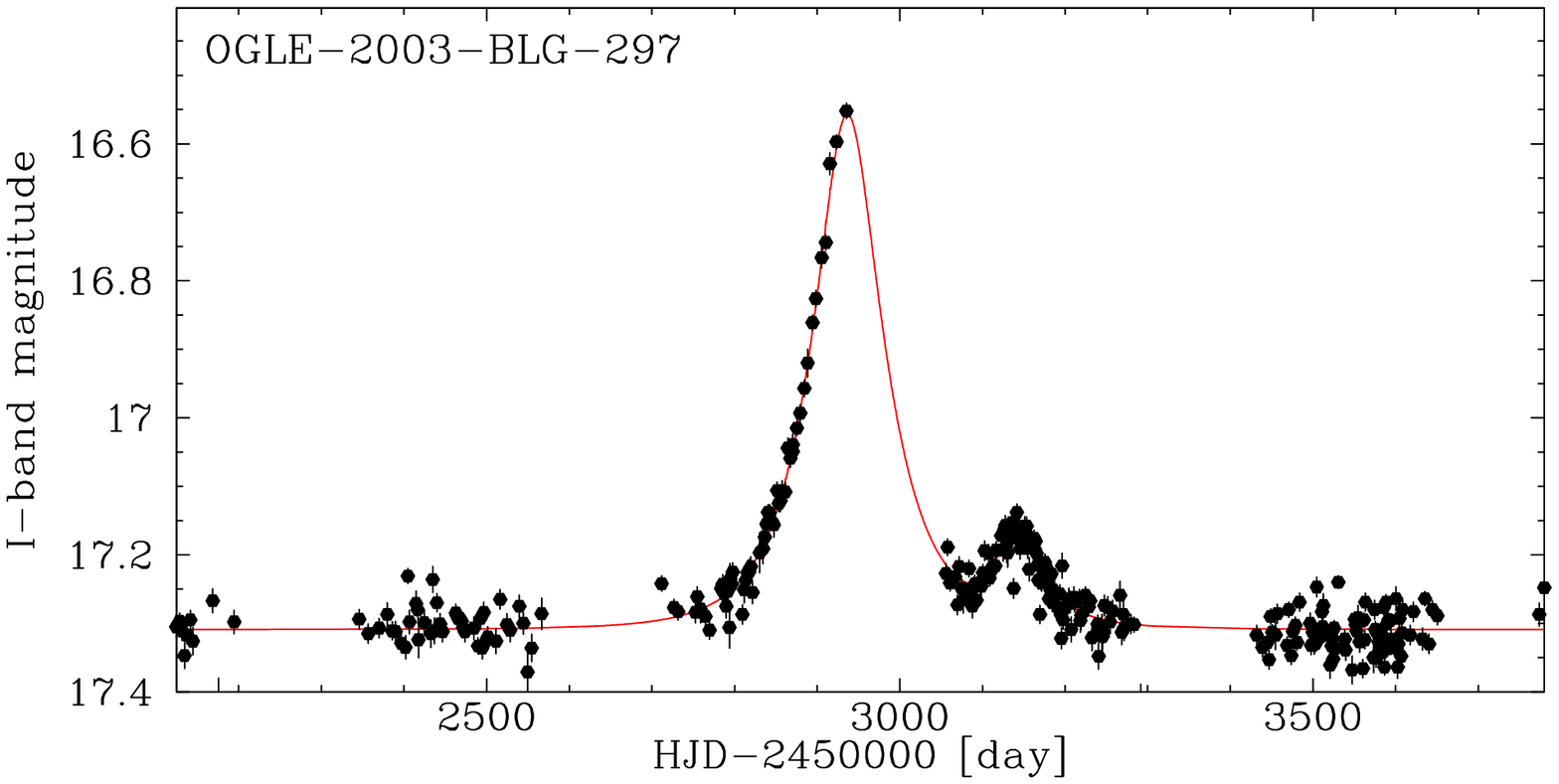}

\noindent
\includegraphics[width=\colwidth]{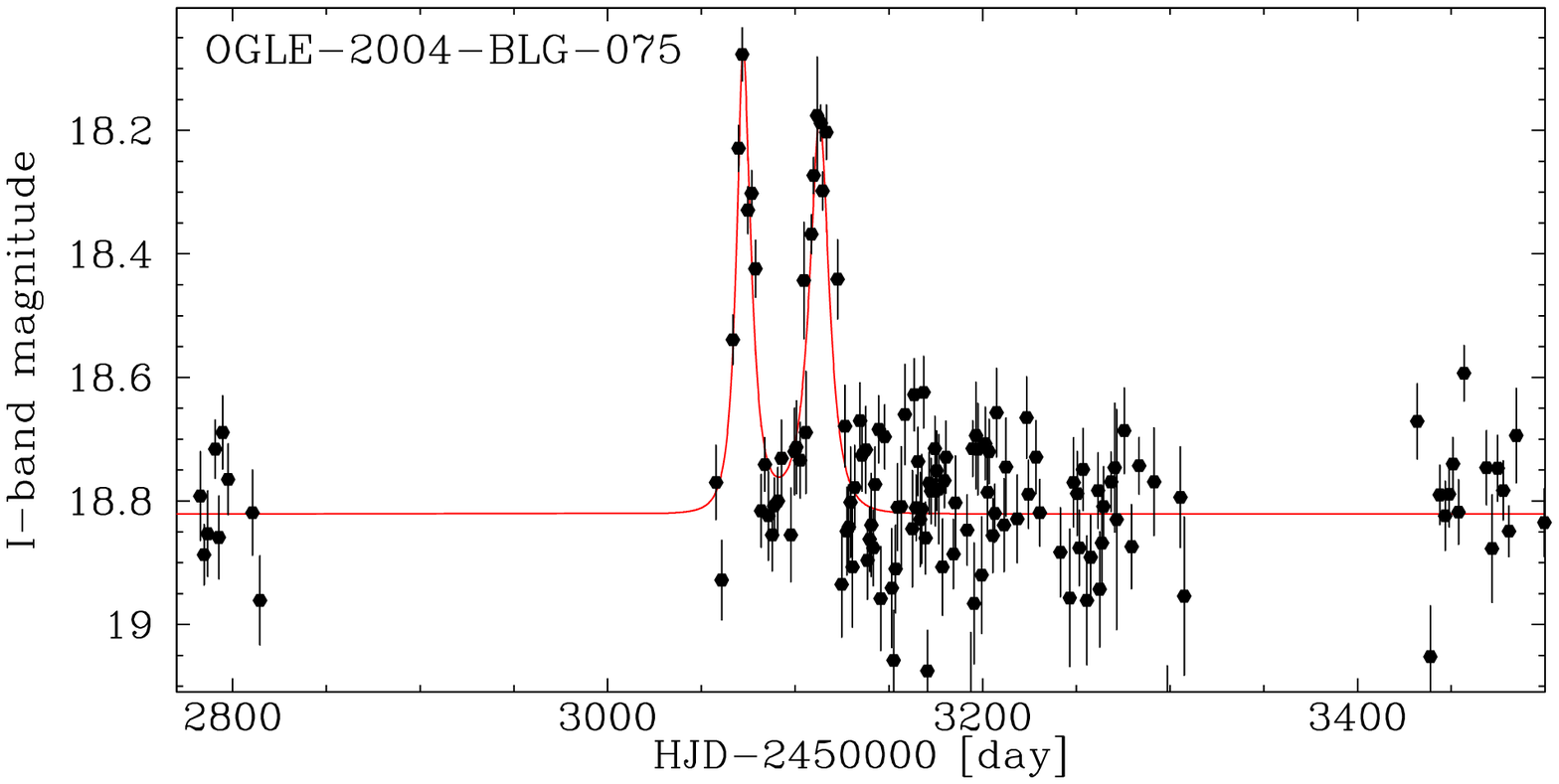}

\noindent
\includegraphics[width=\colwidth]{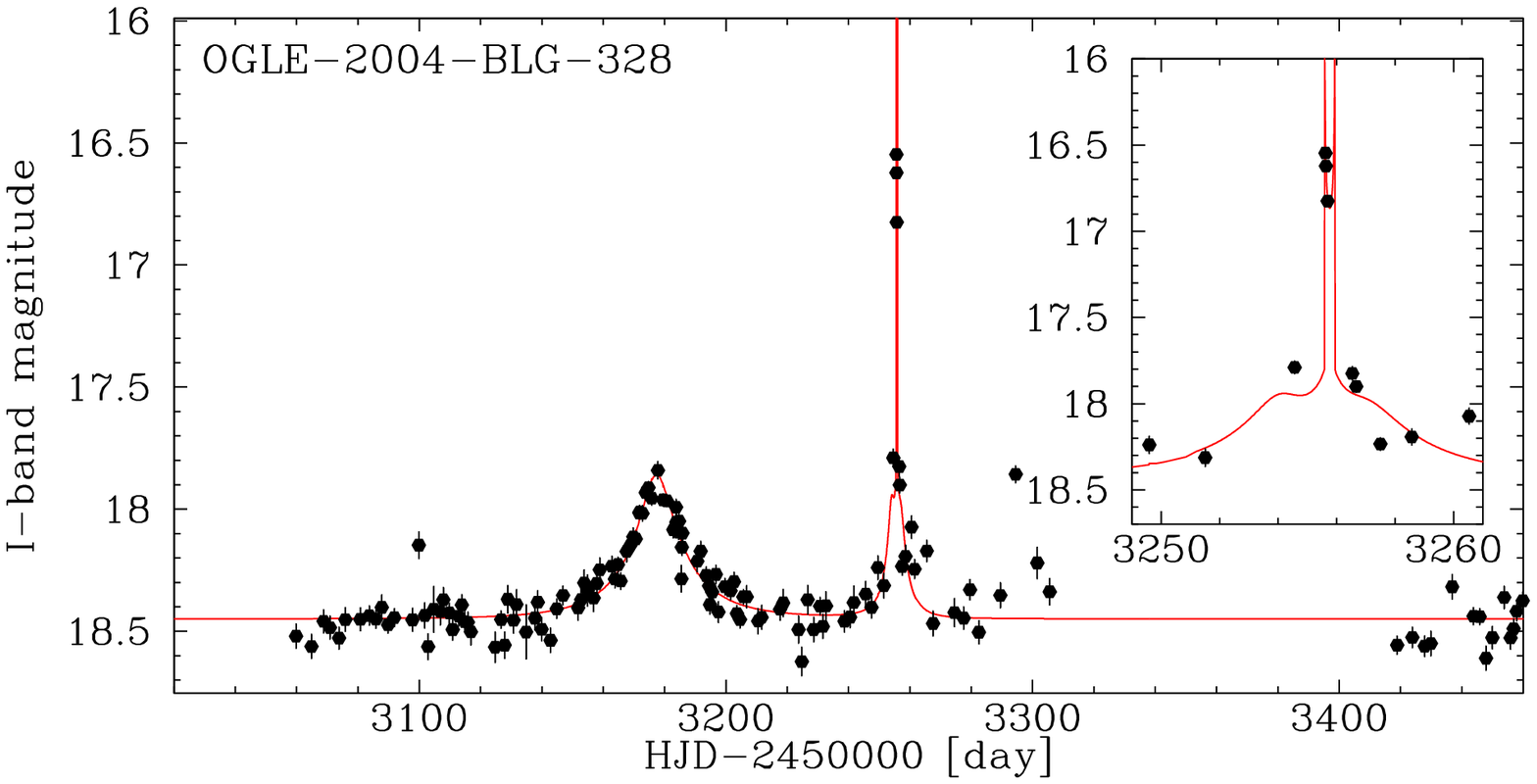}

\noindent
\includegraphics[width=\colwidth]{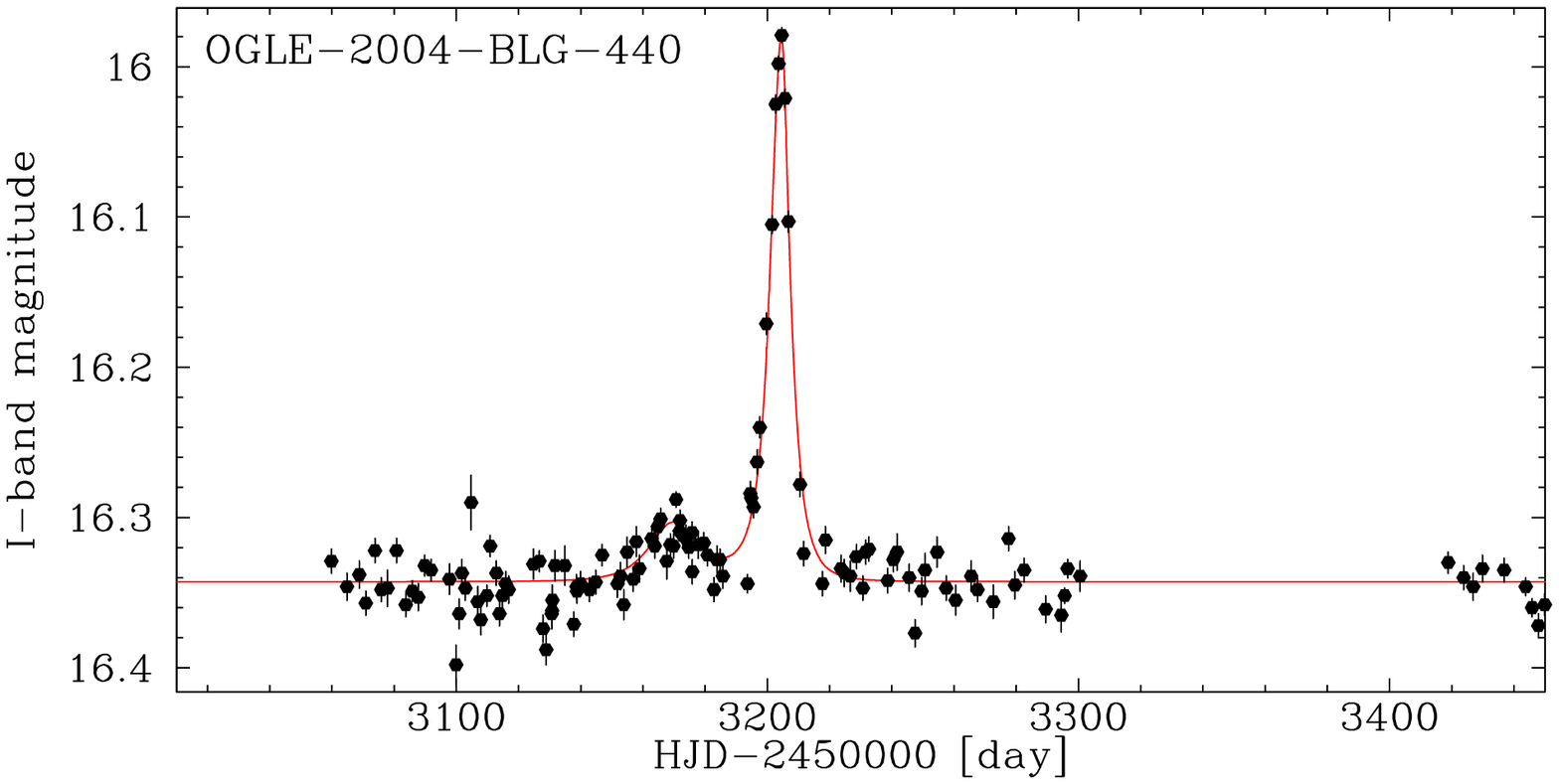}

\noindent
\includegraphics[width=\colwidth]{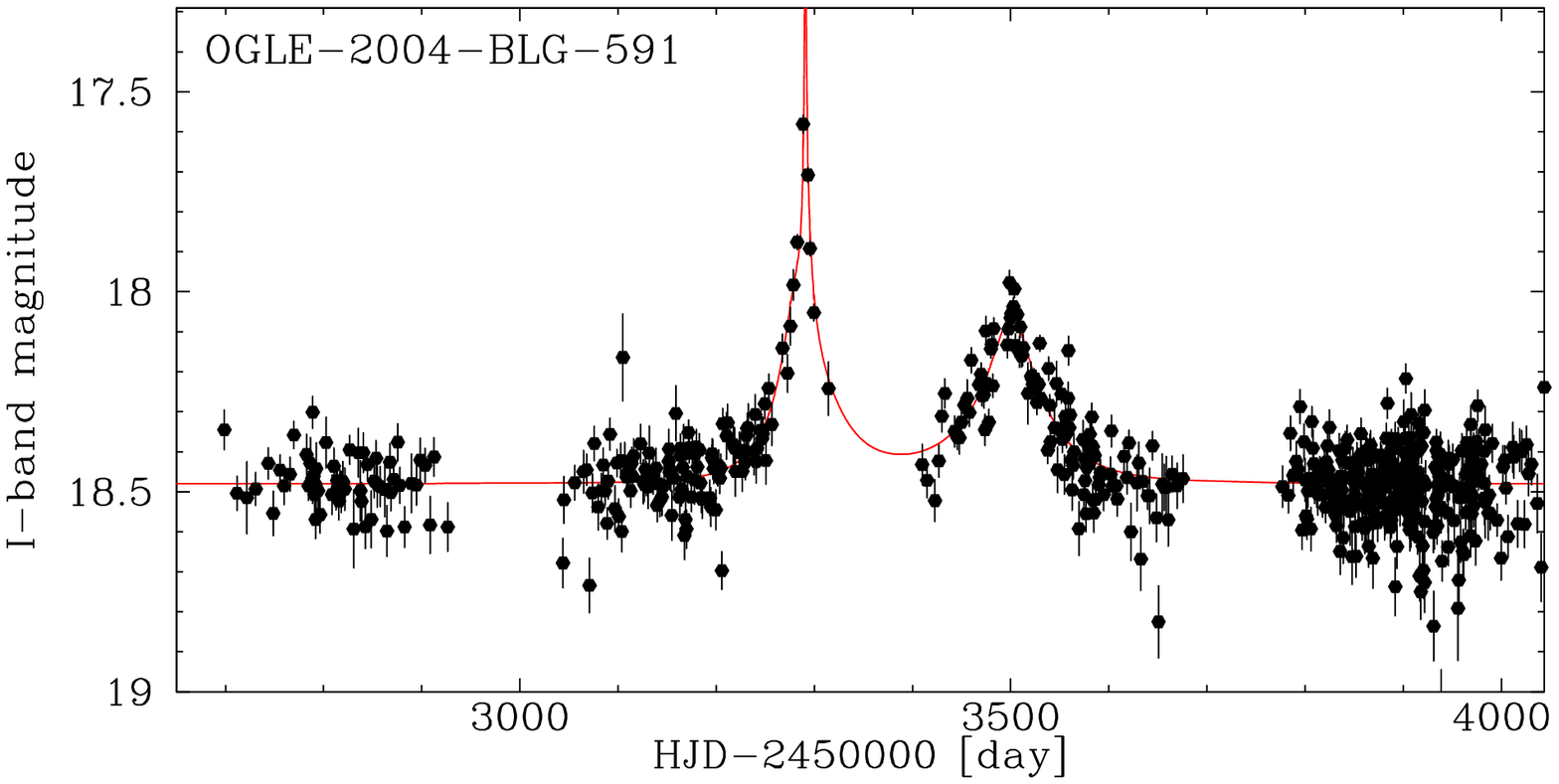}

\noindent
\includegraphics[width=\colwidth]{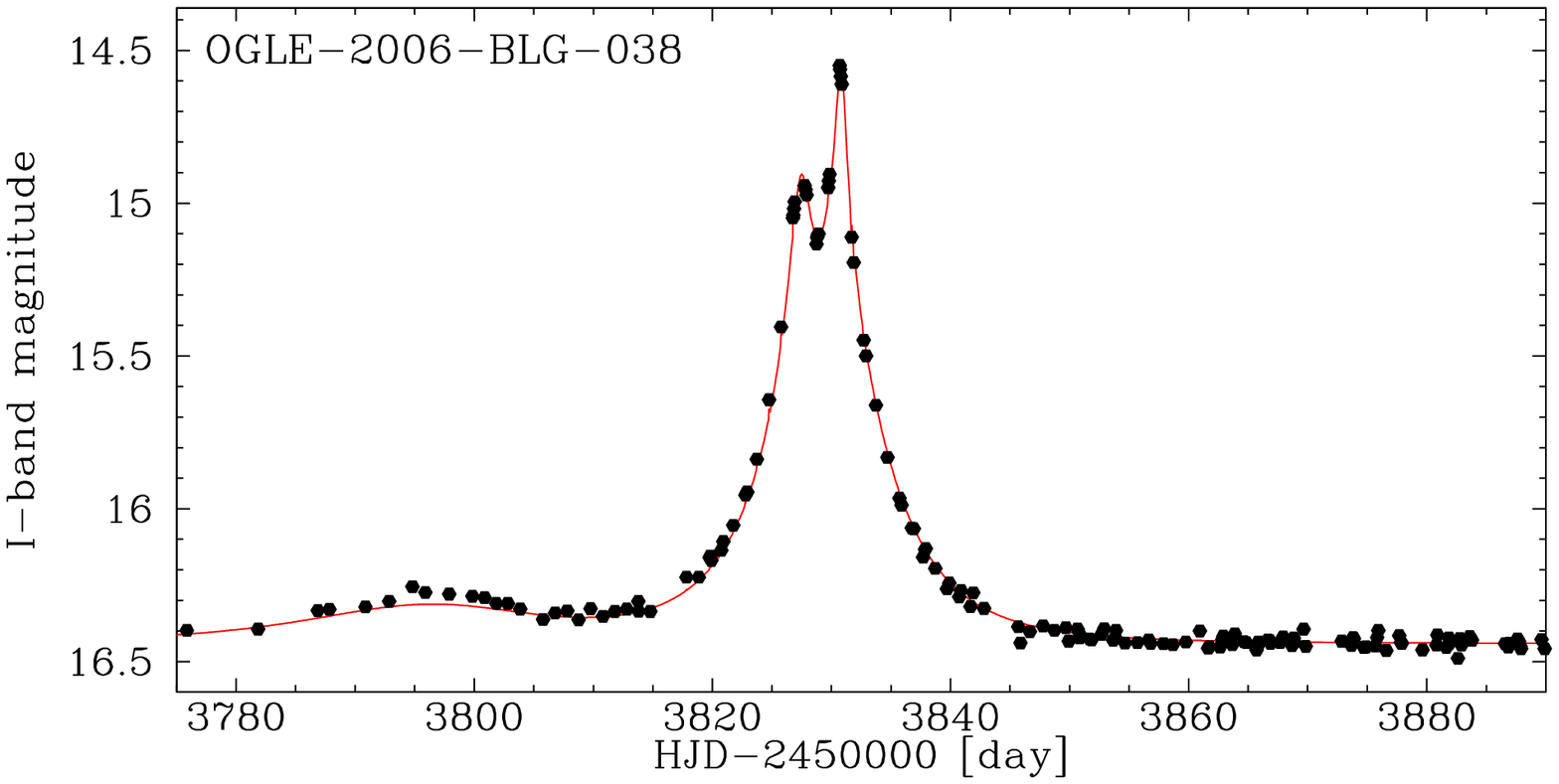}

\noindent
\includegraphics[width=\colwidth]{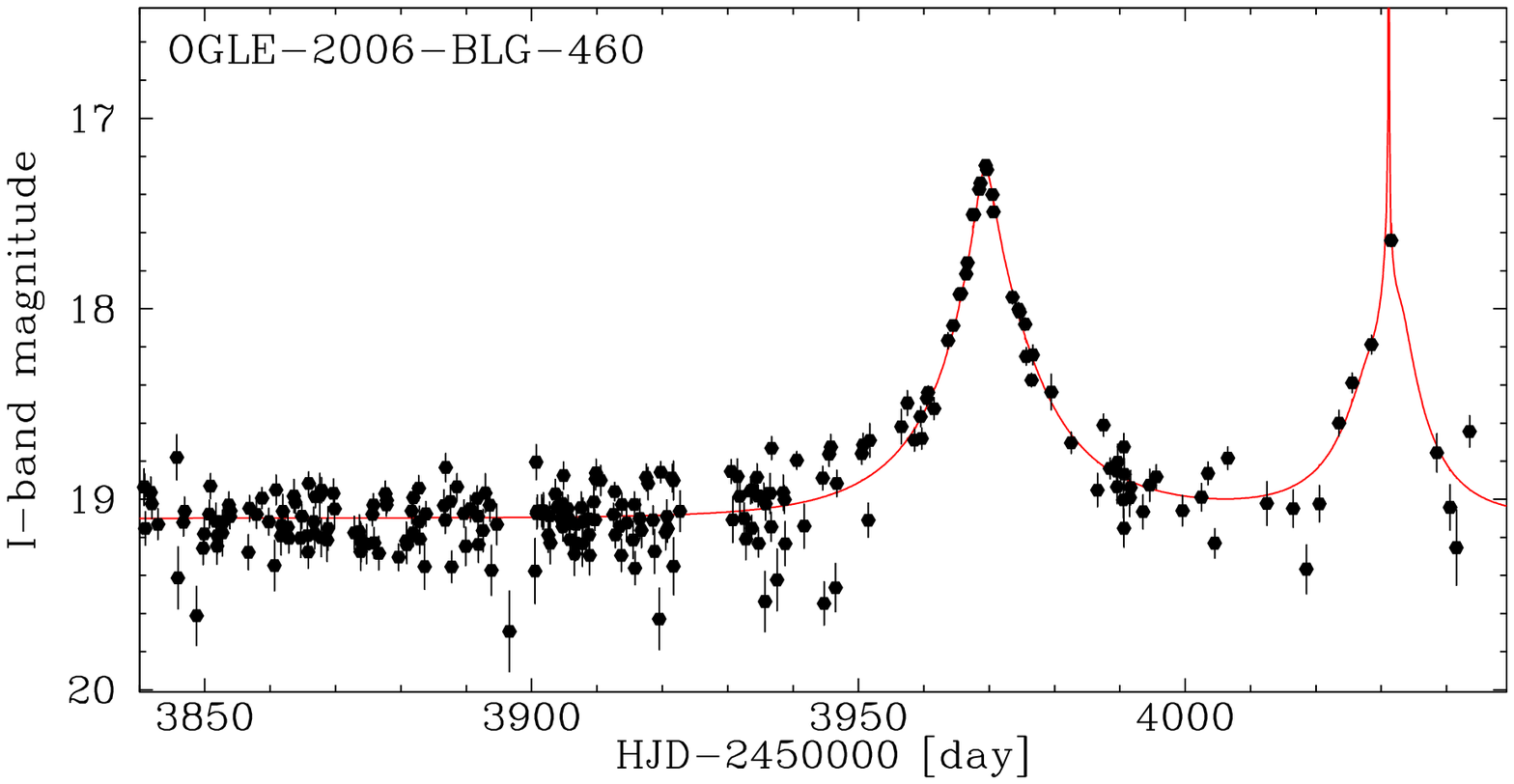}

\noindent
\includegraphics[width=\colwidth]{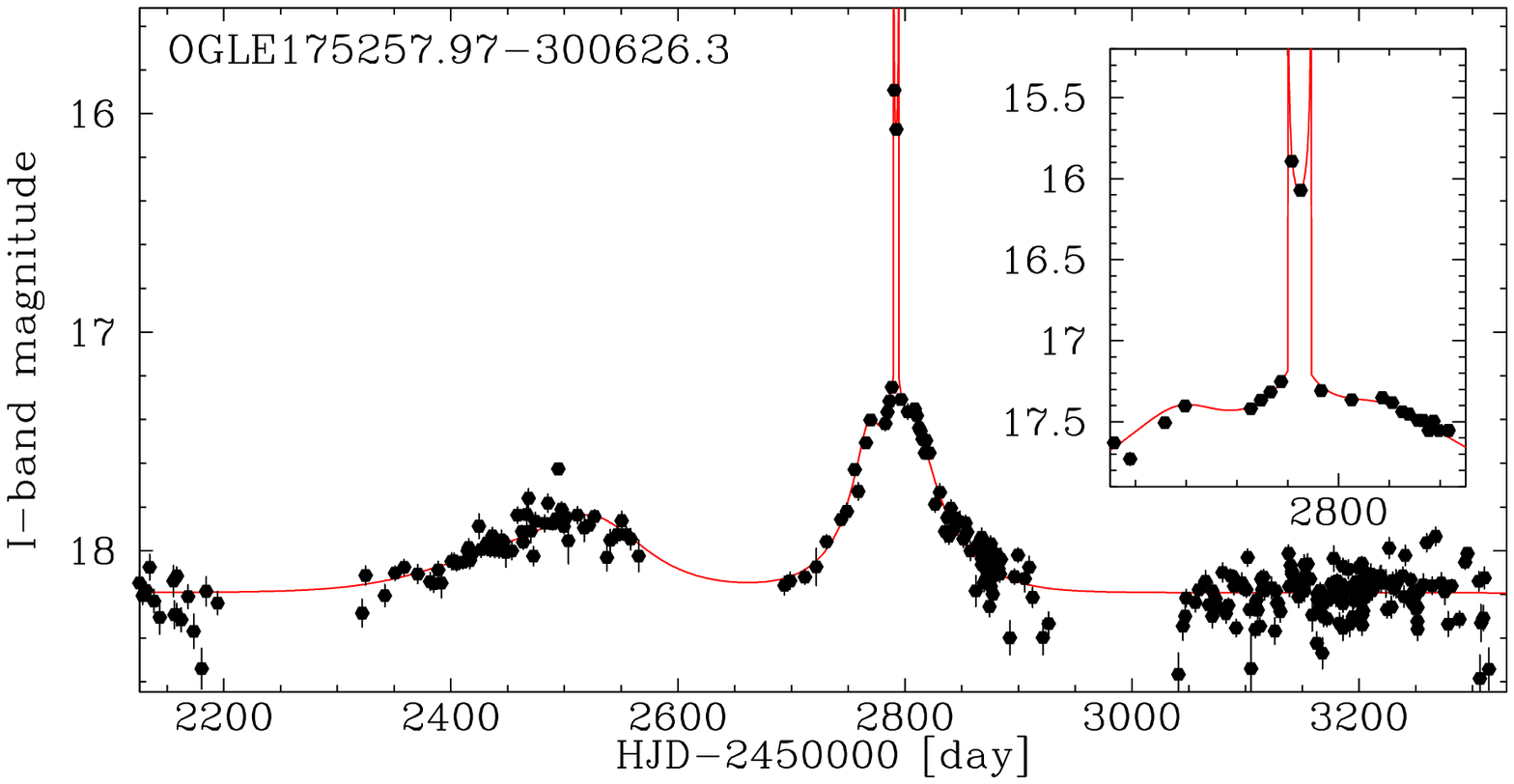}

\noindent
\includegraphics[width=\colwidth]{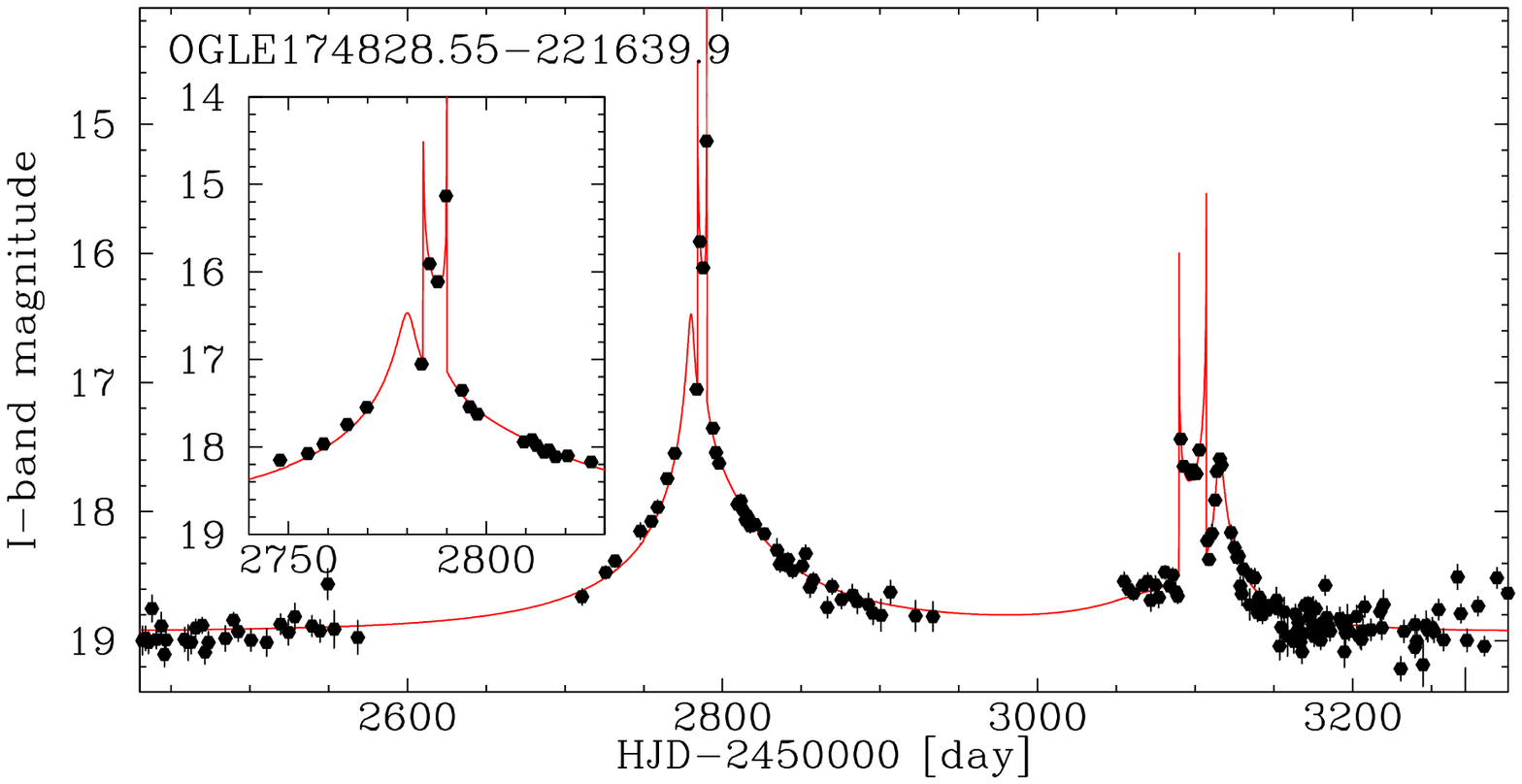}

\end{scriptsize}


\label{lastpage}

\end{document}